\documentclass[aps,prd,twocolumn,amsmath,amssymb,superscriptaddress]{revtex4-1}

\usepackage[utf8]{inputenc}
\usepackage[dvips]{graphicx}
\usepackage{psfrag}
\usepackage{color}

\newcommand{\dd}{\mathrm{d}}
\newcommand{\ee}{\mathrm{e}}

\renewcommand{\vec}[1]{\ensuremath{\boldsymbol{#1}}}

\usepackage{xspace}
\newcommand*{\eg}{e.\,g.\@\xspace}
\newcommand*{\ie}{i.\,e.\@\xspace}
\newcommand*{\cf}{cf.\@\xspace}
\newcommand*{\eq}[1]{Eq.~(\ref{eq:#1})}
\newcommand*{\fig}[1]{Fig.~\ref{fig:#1}}
\renewcommand*{\sec}[1]{Sec.~\ref{sec:#1}}

\newcommand{\itp}{Institut für Theoretische Physik,\\ Universität
Heidelberg, Philosophenweg 16, D--69120 Heidelberg, Germany}
\newcommand{\bologna}{Dipartimento di Fisica e Astronomia, Alma Mater
Studiorum Universit\`a di Bologna, viale Berti Pichat, 6/2, I-40127
Bologna, Italy}
\newcommand{\inaf}{INAF -- Osservatorio Astronomico di Bologna, via
Ranzani 1, I-40127 Bologna, Italy}
\newcommand{\infn}{INFN -- Sezione
di Bologna, viale Berti Pichat 6/2, I-40127 Bologna, Italy}
\newcommand{\kavli}{Institute of Astronomy and Kavli Institute for
Cosmology, University of Cambridge, Madingley Road, Cambridge CB3 0HA,
UK}
\newcommand{\hits}{Heidelberger Institut für Theoretische
Studien, Schloss-Wolfsbrunnenweg 35, 69118 Heidelberg, Germany}

\usepackage{hyperref}
\hypersetup{pdftitle=Nonlinear growig neutrino cosmology}

\begin{document}


\title{Nonlinear growing neutrino cosmology}

\author{Youness Ayaita}
\email[]{ayaita@thphys.uni-heidelberg.de}
\affiliation{\itp}

\author{Marco Baldi}
\affiliation{\bologna}
\affiliation{\inaf}
\affiliation{\infn}

\author{Florian Führer}
\affiliation{\itp}

\author{Ewald Puchwein}
\affiliation{\kavli}
\affiliation{\hits}

\author{Christof Wetterich}
\affiliation{\itp}

\date{\today}

\begin{abstract}
	The energy scale of Dark Energy, $\sim 2\times 10^{-3}$~eV, is a
	long way off compared to all known fundamental scales~-- except
	for the neutrino masses. If Dark Energy is dynamical and couples
	to neutrinos, this is no longer a coincidence. The time at which
	Dark Energy starts to behave as an effective cosmological constant
	can be linked to the time at which the cosmic neutrinos become
	nonrelativistic. This naturally places the onset of the Universe's
	accelerated expansion in recent cosmic history, addressing the
	{\it why-now} problem of Dark Energy. We show that these
	mechanisms indeed work in the {\it Growing Neutrino Quintessence}
	model~-- even if the fully nonlinear structure formation and
	backreaction are taken into account, which were previously
	suspected of spoiling the cosmological evolution. The attractive
	force between neutrinos arising from their coupling to Dark Energy
	grows as large as $10^6$ times the gravitational strength. This
	induces very rapid dynamics of neutrino fluctuations which are
	nonlinear at redshift $z \approx 2$. Nevertheless, a nonlinear
	stabilization phenomenon ensures only mildly nonlinear oscillating
	neutrino overdensities with a large-scale gravitational potential
	substantially smaller than that of cold dark matter perturbations.
	Depending on model parameters, the signals of large-scale neutrino
	lumps may render the cosmic neutrino background observable.
\end{abstract}


\maketitle



\section{Introduction}
\label{sec:introduction}

The cosmological constant $\Lambda$ has emerged as the standard
explanation for the observed accelerated expansion of the Universe
\cite{Perlmutter99, Riess98}. Together with the assumption of cold
dark matter (CDM), it forms the remarkably successful concordance
model $\Lambda$CDM \cite{Bartelmann10}. The proposed alternatives to
the cosmological constant are already many~-- more complicated and
often a worse fit to observational data \cite{Copeland06}. A new model
should only be added to this list if it provides theoretical
advantages or phenomenological aspects that neither the cosmological
constant nor its most prominent competitors can offer. Growing
Neutrino Quintessence (GNQ) was proposed in this spirit
\cite{Amendola08, Wetterich07}. It addresses both the cosmological
constant problem (why is the energy density of Dark Energy so small?)
and the why-now problem (why has Dark Energy just started to dominate
the energy budget of the Universe?) \cite{Weinberg89, Carroll01}. On
the phenomenological side, it predicts a time-varying neutrino mass
and the formation of large-scale neutrino overdensities that might be
detectable by their gravitational potentials \cite{Mota08}.

As a quintessence model \cite{Wetterich88, Ratra88}, GNQ describes the
Dark Energy by a dynamical scalar field, the cosmon $\varphi$.
Analogously to the inflaton in inflationary theories of the early
Universe, the cosmon can describe an accelerated expansion of the
Universe at late times. The similarity of the mechanism even allows
for a unified picture in which the same field is responsible for both
the early and the late accelerating epochs \cite{Wetterich13,
Wetterich14}. Quintessence models address the cosmological constant
problem: the energy density of Dark Energy decays, during most of the
cosmological evolution, just like that of radiation and matter. Its
small size today is then simply a consequence of the large age of the
Universe.

In contrast to the simplest quintessence models, GNQ includes a
mechanism for a natural crossover to the accelerated phase. No
fine-tuning of the self-interaction potential is needed. Instead, a
coupling between the cosmon and the neutrinos affects the dynamics of
Dark Energy. The event of the cosmic neutrinos becoming
nonrelativistic --~which, due to their small masses, happens in
relatively recent cosmic history~-- triggers the onset of Dark Energy
domination.

This requires a relatively strong coupling between the cosmon and the
neutrinos, which can have a natural explanation in a particle physics
setting \cite{Wetterich07}. Such a coupling has, however, a decisive
impact on the evolution of perturbations in the neutrino density. The
perturbations become nonlinear even on very large scales
\cite{Mota08}. Furthermore, the expansion history can be affected by a
nonlinear backreaction effect \cite{Pettorino10}. These technical
complications motivated a comprehensive simulation technique
\cite{Ayaita12a}. The technique has by now matured and allows to
obtain full cosmological evolutions of the model. In the technically
simpler case of a constant coupling parameter, its preliminary results
already inspired a consistent physical picture and an approximation
scheme for the nonlinear evolution \cite{Ayaita12b}. In this work, we
will turn to the more natural yet technically challenging case of a
field-dependent coupling. Again, a coherent (though fundamentally
different) physical picture of the cosmological evolution will emerge.
Our results for the first time show the full cosmological evolution of
GNQ until redshift zero.

\begin{figure}
	\begin{center}
		\psfrag{xlabel}[B][c][.9][0]{scale factor $a$}
		\psfrag{ylabel}[B][c][.9][0]{energy fractions $\Omega_i$}
		\includegraphics[width=0.45\textwidth]{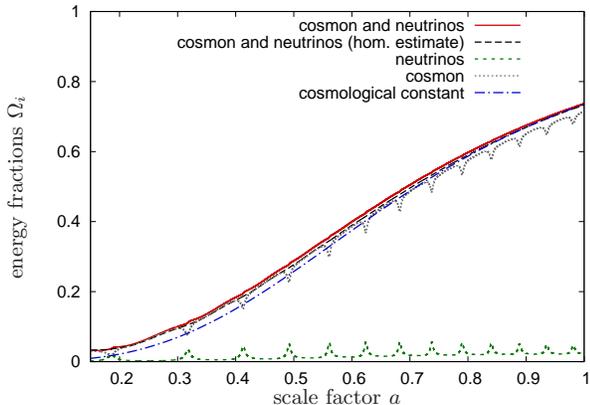}
	\end{center}
	\caption{The transition to Dark Energy domination in nonlinear
	growing neutrino cosmology. The figure shows the energy fraction
	of the coupled cosmon-neutrino fluid as obtained in a nonlinear
	simulation (red solid) and a purely homogeneous computation
	(black dashed). The two lines almost coincide, demonstrating the
	smallness of the ``backreaction''. The individual components are
	the cosmon (gray dotted) and the neutrinos (green dotted). The
	cosmological evolution of the Dark Energy fraction is compared to
	a cosmological constant normalized to the same present-day Dark
	Energy density (blue dot-dashed).}
	\label{fig:omegas}
\end{figure}

A relation between Dark Energy (in the form of a scalar field) and the
neutrino masses has earlier been studied in models of `mass-varying
neutrinos' (MaVaNs) \cite{Fardon03}. These models share certain
features with GNQ, in particular the instability problem of neutrino
perturbations \cite{Afshordi05, Bjaelde07, Brouzakis07}. The
cosmon-neutrino coupling, once strong enough, can lead to the
formation of large nonlinear neutrino lumps. These lumps would, as a
backreaction effect, influence the expansion dynamics of the Universe.
They could even prevent the Universe from entering a phase of
accelerated expansion. For GNQ, the strong backreaction effect of
stable neutrino lumps on the expansion dynamics has been shown in a
simulation \cite{Ayaita12a}. Our results, however, provide a
counterexample in which --~in spite of the instability of
perturbations~-- the backreaction effect remains small and the
expansion dynamics is affected only marginally. We anticipate this
numerical result in \fig{omegas}. 
Although we will encounter sizable backreaction effects in the
interaction of Dark Energy and neutrinos, the backreaction effect on
the combined cosmon-neutrino fluid is hardly visible. The evolution of
the energy density of this fluid is very similar to that of a
cosmological constant. The main distinction is the presence of a small
fraction of Early Dark Energy.

This work is organized as follows. The next section covers a brief
overview of the fundamentals of the model and the most important
insights into its cosmological evolution that preceded this work.
Section~\ref{sec:method} will explain the main ideas of the simulation
method. The numerical results in Sec.~\ref{sec:results} are followed
by a physical interpretation in Sec.~\ref{sec:interpretation}
sketching a coherent physical picture of the evolution. The work
concludes in Sec.~\ref{sec:conclusion}.


\section{Growing Neutrino Quintessence}
\label{sec:qng}

\subsection{Basic concepts}
\label{sec:basics}

In this section, we briefly collect and explain the main ingredients
that make up the GNQ model. In a nutshell, these are the cosmon
$\varphi$ described as a scalar field with a canonical kinetic term
and a self-interaction potential $V(\varphi)$ and the neutrinos whose
masses are assumed to depend on $\varphi$. The field-dependence of the
neutrino mass defines an interaction between the cosmon and the
neutrinos whose coupling parameter $\beta$ is a measure of how strong
this field-dependence is.

Let us take the time to go through this in more detail. The Lagrangian
of the cosmon alone is of standard form
\begin{equation}
	-\mathcal L_\varphi = \frac{1}{2} \partial^\lambda \varphi
	\partial_\lambda \varphi + V(\varphi).
	\label{}
\end{equation}
Here and in the following, we use the metric signature $(-,+,+,+)$ and
units where the reduced Planck mass is unity, implying $8 \pi G = 1$.
We assume an exponential potential $V(\varphi) \propto \exp(-\alpha
\varphi)$ \cite{Wetterich08}. The details of the potential do not
matter as long as it gives rise to suitable scaling solutions ensuring
--~for a wide range of initial conditions~-- that Dark Energy decays
just as the dominant component (radiation and later matter). In our
case, the constraints on Early Dark Energy require $\alpha \gtrsim 10$
\cite{Doran07, Reichardt12, Pettorino13}. The scaling solution should
hold as long as neutrino masses play no role.

The second ingredient is the dependence of the neutrino masses on
$\varphi$ \cite{Wetterich07}. For simplicity, we only consider the
average neutrino mass $m_\nu$ instead of the full mass matrix $M_\nu$
of the three light neutrinos. A dependence $m_\nu = m_\nu(\varphi)$
occurs if a fundamental mass scale $M$ in the mechanism of neutrino
mass generation depends on $\varphi$. For example, in the {\it
cascade} or {\it induced triplet} mechanism \cite{Magg80, Lazarides81,
Mohapatra80, Schechter80}, the neutrino masses are proportional to
$M_t^{-2}$ where $M_t$ denotes the mass of a heavy $\text{SU}(2)_L$
triplet. If $M_t$ depends on $\varphi$ such that it reaches a small
value near $\varphi = \varphi_\text{crit}$, the average neutrino mass
can be approximated, in the range of interest, by the ansatz
\begin{equation}
	m_\nu(\varphi) = \frac{\bar m}{\varphi_\text{crit} - \varphi}
	\label{eq:divmass}
\end{equation}
with a parameter $\bar m$ \cite{Wetterich07}. The formal pole at
$\varphi_\text{crit}$ is never reached by the cosmological solution
and may be considered as an artifact of the approximation. Also the
behavior far away from this pole is not important for our
considerations as, in this case, the cosmon-neutrino coupling is
negligible. We can thus employ the relation given by \eq{divmass} for
the full cosmological evolution.

The cosmon-neutrino coupling $\beta$ quantifies the strength of the
field-dependence of $m_\nu$. It is defined as
\begin{equation}
	\beta(\varphi) \equiv -\frac{\dd \ln m_\nu(\varphi)}{\dd\varphi}
	= - \frac{1}{\varphi_\text{crit} - \varphi},
	\label{eq:divbeta}
\end{equation}
where, in the last step, we have used the explicit dependence of
\eq{divmass}. When $\varphi$ approaches $\varphi_\text{crit}$, the
coupling becomes strong and successfully stops the evolution of the
cosmon. However, other functional shapes for $\beta$ are possible as
well. For instance, a technically simple choice is a constant coupling
$\beta = \text{const.}$ implying an exponential mass dependence
$m_\nu(\varphi) \propto \exp(-\beta\varphi)$. A growing neutrino mass
requires a negative coupling parameter $\beta < 0$. The coupling
between the cosmon and the neutrinos manifests itself as an
energy-momentum exchange between the two components. This
energy-momentum transfer is proportional to the coupling parameter
$\beta(\varphi)$ and reads
\begin{align}
	\nabla_\lambda T^{\mu\lambda}_{(\varphi)} &= + \beta(\varphi)\,
	T_{(\nu)} \partial^\mu \varphi, \label{eq:exch1} \\
	\nabla_\lambda T^{\mu\lambda}_{(\nu)} &= - \beta(\varphi)\,
	T_{(\nu)} \partial^\mu \varphi, \label{eq:exch2}
\end{align}
where $T_{(\nu)} \equiv T^\lambda_{(\nu)\,\lambda} = -\rho_\nu + 3
p_\nu$ denotes the trace of the neutrino energy-momentum tensor.
Quintessence couplings of this simple type are discussed in early
works on coupled Dark Energy \cite{Wetterich95, Amendola00}.

Inserting the cosmon's energy-momentum tensor in \eq{exch1} yields the
field equation
\begin{equation}
	\nabla^\lambda \nabla_\lambda \varphi - V_{,\varphi}(\varphi)
	= \beta(\varphi)\, T_{(\nu)}.
	\label{eq:kg}
\end{equation}
It shows that the cosmon-neutrino coupling becomes only effective once
the right-hand side, $\beta T_{(\nu)}$, is comparable to or larger
than the potential derivative $V_{,\varphi}$. As long as the neutrinos
are relativistic with $w_\nu \approx 1/3$, the trace $T_{(\nu)} = -
\rho_\nu(1 - 3 w_\nu)$ and thereby the effect of the coupling is
negligible. In this sense, the neutrinos becoming nonrelativistic
serves as a trigger event. On the other hand, also $\beta(\varphi)$
grows towards large negative values as $\varphi$ rolls down its
potential towards $\varphi_\text{crit}$. This ensures that,
eventually, the effect of the coupling cancels the effect of the
potential derivative. In that case, the evolution of the cosmon is
essentially stopped, and the Dark Energy approximately acts as a
cosmological constant with vacuum energy $V(\varphi_\text{crit})$. We
will find that $\varphi$, and therefore the neutrino masses, oscillate
around a slowly increasing value.

For a neutrino particle on a classical path, the coupling implies the
equation of motion \cite{Ayaita12a}
\begin{equation}
	\frac{\dd u^\mu}{\dd \tau} + \Gamma^\mu_{\alpha\beta} u^\alpha
	u^\beta = \beta(\varphi)\, \partial^\mu \varphi
	+ \beta(\varphi)\, u^\lambda \partial_\lambda \varphi \, u^\mu,
	\label{eq:eom}
\end{equation}
where $u^\mu$ is the four-velocity and $\tau$ denotes the proper time.
The left-hand side is simply the motion under gravity, whereas the
right-hand side includes the effects of the cosmon-neutrino coupling.
For the (spatial) velocities $u^k$, the first term, $\beta \partial^k
\varphi$, is similar to a potential gradient in Newtonian gravity and
can be interpreted as an attractive force between the neutrinos. In
the limit of small velocities, it is about $2\beta^2$ stronger than
gravity \cite{Wintergerst10a}. For relativistic velocities, it becomes
negligible as the other contributions grow quadratically with
components of the four-velocity $u^\mu$. In this case, the coupling is
only important in the second term on the right-hand side, which,
however, cannot change the direction of motion of the particle. Thus,
the cosmon-mediated attraction of neutrinos is only effective in the
nonrelativistic case.

A second important ingredient is the replacement of the Hubble damping
by ``cosmon acceleration''. Neglecting the (spatial) gradients
$\partial^k \varphi$ (and, similar, for the metric), \eq{eom} becomes
($u^0 = \gamma$)
\begin{align}
	\frac{\dd u^k}{\dd t} &= \left[ \beta(\varphi)\dot\varphi - 2 H
	\right]\, u^k,
	\label{eq:cosmonacc1} \\
	\frac{\dd \gamma}{\dd t} &= \left[ \beta(\varphi)\dot\varphi - H
	\right]\,\frac{\gamma^2 - 1}{\gamma}.
	\label{eq:cosmonacc2}
\end{align}
(This is consistent with the defining relation $\gamma^2 = 1 + a^2 u^k
u_k$.) For an expanding universe, the positive sign of $H$ induces a
damping of all motions. We will find that the contribution $\propto
\beta\dot\varphi$ overwhelms the Hubble damping for important periods
in the formation of nonlinear neutrino structures. The acceleration of
all neutrino motions for $\dot\varphi < 0$ will play a crucial role
for the dissolution of previously formed neutrino lumps.

\subsection{Cosmon-neutrino structure formation for constant coupling}
\label{sec:structure}

Understanding structure formation in GNQ is not only important to make
contact with various observational constraints such as from the cosmic
microwave background (CMB) or galaxy surveys. It even is a
prerequisite for obtaining reliable estimates of the expansion
dynamics. This is because, as we will review in this section,
nonlinear perturbations in the cosmon-neutrino fluid can lead to
strong backreaction effects. They alter cosmological averages of the
neutrino mass and equation of state, which, in turn, influences the
evolution of Dark Energy at the background level. We explain this by
briefly reviewing the main steps undertaken by previous works that
have shed light on the issue \cite{Brouzakis07, Mota08, Schrempp09,
Wintergerst10a, Pettorino10, Nunes11, Baldi11a, Ayaita12a, Ayaita12b}.
These works focused on the constant coupling model where $\beta$ does
not depend on $\varphi$. It is technically simpler and may be regarded
as a useful approximation in the case where $\beta(\varphi)$ does not
vary much in late cosmology. Obtaining a realistic accelerated
expansion requires couplings of order $\beta \sim -10^2$ if the
potential is exponential with $\alpha \gtrsim 10$.

\begin{figure}[htb!]
	\begin{center}
		\psfrag{xlabel}[B][c][.9][0]{comoving scale $k$ [$h/$Mpc]}
		\psfrag{ylabel}[B][c][.9][0]{transitional scale factor
		$a_\text{nl}$}
		\includegraphics[width=0.45\textwidth]{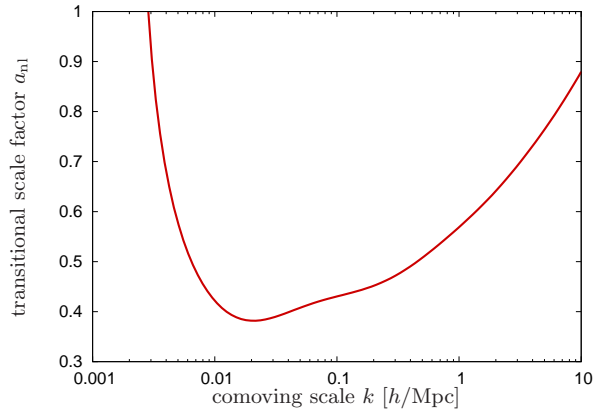}
	\end{center}
	\caption{Onset of nonlinearity for neutrino fluctuations. We show
	the scale factor $a_\text{nl}$ at which different scales enter the
	nonlinear regime. Compare with the corresponding figure in
	Ref.~\cite{Mota08}.}
	\label{fig:anl}
\end{figure}

The large value of the coupling implies a fast growth of linear
neutrino perturbations. The transition to the nonlinear regime can be
associated roughly with the moment at which the dimensionless power
spectrum $\Delta_\nu^2(k) = k^3 P_\nu(k)/(2\pi^2)$ reaches order
unity. In contrast to the CDM case, this transition occurs even on
very large scales leading to a breakdown of linear perturbation theory
\cite{Mota08}. Let $a_\text{nl}(k)$ denote the cosmic scale factor at
which $\Delta_\nu^2(k) = 1$ in linear perturbation theory.
Figure~\ref{fig:anl} shows the transition to nonlinearity (in the
Newtonian gauge) for $\beta = -52$, $\alpha = 10$, and a relatively
large present-day average neutrino mass $m_\nu^0 \approx 2.3$~eV.
Although the details depend on the precise parameters chosen, the
qualitative finding is generic and has its origin in the instability
of linear perturbations (\cf \sec{basics}).

The first scales to enter the nonlinear regime are of comoving size
$\lambda \sim 100 h^{-1}$Mpc (\cf \fig{anl}). The overdensities at
this scale evolve into massive neutrino lumps that are stable for
constant $\beta$. Although the picture will be different for the
varying (\ie cosmon-dependent) coupling $\beta(\varphi)$ investigated
in this work, \cf \eq{divbeta}, it is worthwhile to discuss the main
effects in the technically simpler setting of a constant coupling
$\beta \sim -10^2$. They will reappear, albeit in a weaker form, in
the varying coupling model and play a role in the physical
interpretation of our results.

Two properties of the lumps were identified that imply a backreaction
effect altering the expansion dynamics \cite{Nunes11}. They both lead
to a suppression
\begin{equation}
	| \overline{ T_{(\nu)} } | < | T_{(\nu)}^\text{hom} |
	\label{}
\end{equation}
of the actually averaged trace of the neutrino energy-momentum tensor
as compared to the trace obtained in a purely homogeneous computation
that neglects the nonlinear perturbations. It is, however, this trace
$T_{(\nu)}$ that enters the cosmon field equation, \eq{kg}. The more
severely the trace is suppressed, the less effective is the coupling
in stopping the evolution of the cosmon.

\begin{figure}[htb!]
	\begin{center}
		\scalebox{0.65}{\input{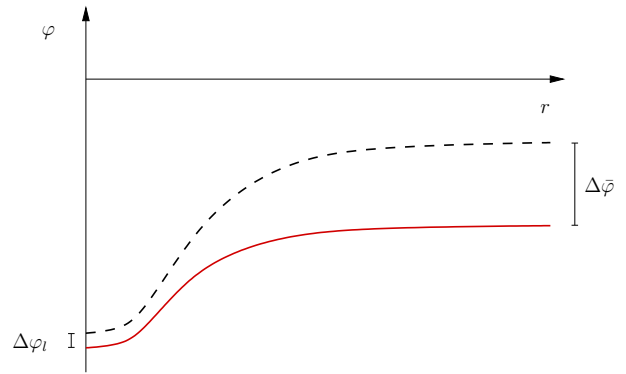}}
	\end{center}
	\caption{The local value of the cosmon effectively decouples from
	the background evolution leading to an approximate mass freezing
	within lumps. We use arbitrary units in this schematic plot, where
	the core of the potential well is typically of the size of several
	Megaparsecs and the depth of the potential well is of order $\bar
	\varphi - \varphi(0) \sim 10^{-2}$ \cite{Ayaita12a}.}
	\label{fig:backreaction}
\end{figure}

First, during the lump formation, the neutrinos are accelerated to
higher velocities. This can lead, in particular close to the lumps'
centers, to relativistic neutrino velocities. Those neutrinos no
longer contribute to $\overline{ T_{(\nu)} }$ as the energy-momentum
tensor of relativistic particles is approximately traceless. Second,
similarly to a gravitational potential well, the local cosmon
perturbation $\delta\varphi$ is negative in lumps, leading to neutrino
masses $m_\nu(\bar\varphi + \delta\varphi)$ that are smaller inside
the lumps than expected for the cosmological average field
$\bar\varphi$. As $T_{(\nu)} \propto m_\nu$, this substantially
weakens the effect of the coupling as most neutrinos will be located
in lumps. The effect can be physically understood as an approximate
mass freezing within lumps~-- the nonlinear lumps approximately
decouple from the background; the local value $\varphi_l$ of the
cosmon within the lumps no longer follows the evolution of the
homogeneous component $\bar \varphi$ \cite{Nunes11}. We illustrate
this schematically in \fig{backreaction}.
In this illustration, a stable lump is located at $r = 0$; within some
time interval $\Delta t$, the cosmon evolves by $\Delta \varphi_l$
within and by $\Delta \bar \varphi$ far outside the lump. The figure
tells us that $\varphi_l < \bar\varphi$ and $\Delta \varphi_l < \Delta
\bar \varphi$. Stated differently, the background component
$\bar\varphi$ feels a smaller neutrino mass $\overline{ m_\nu(\varphi)
} < m_\nu(\bar \varphi)$, which, in addition, depends more weakly on
$\bar\varphi$ \cite{Pettorino10}. This weaker dependence can be
expressed as a weaker effective coupling
\begin{equation}
	\beta_\text{eff} \equiv - \frac{\dd \ln
	\overline{m_\nu(\varphi)}}{\dd \bar \varphi},\ |\beta_\text{eff}|
	< |\beta|.
	\label{}
\end{equation}
It postpones the onset of the accelerated expansion \cite{Ayaita12a}.

For constant $\beta$, a clear physical picture and a resulting
approximation scheme have emerged that describe the cosmological
evolution after the formation of lumps \cite{Ayaita12b}. Despite
relativistic neutrino velocities in the lumps' cores, the total lumps
--~also including the local cosmon perturbation~-- behave like
nonrelativistic objects. This is because the negative pressure
contribution of the local cosmon perturbation just cancels the
positive neutrino pressure. The mutual attraction between these
cosmon-neutrino lumps and the interaction between the nonrelativistic
cosmon-neutrino lump fluid and the background cosmon $\bar\varphi$ are
governed by effective couplings weaker than the fundamental coupling
$\beta$.

In this work, we investigate a field-dependent $\beta$ given by
\eq{divbeta}. We will find that important features behave
qualitatively different since lumps turn out to be no longer stable.


\section{Method}
\label{sec:method}

\subsection{An $N$-body approach}
\label{sec:nbody}

Usually, $N$-body simulations are employed to understand the nonlinear
small-scale dynamics of a cosmological model whereas the evolution of
the homogeneous background and of large-scale linear perturbations can
be obtained by simpler means. Not so in GNQ: The effects of nonlinear
perturbations have an impact on all scales including the homogeneous
background (\cf \sec{structure}). As a consequence, a nonlinear method
is bitterly needed in order to understand the cosmological dynamics of
the model.

The first step towards an $N$-body simulation of GNQ was to
incorporate the cosmon-mediated attraction between neutrinos in the
Newtonian limit \cite{Baldi11a}. In this setting, the attractive force
is analogous to gravity but stronger by a factor $2 \beta^2$. The
simulation was capable of describing the first stages of the nonlinear
evolution in which large neutrino lumps started to form. However, the
simplifying assumptions of the approach subsequently broke down.
First, the approach is only valid for nonrelativistic neutrino
velocities; but the neutrinos reached, due to the attractive force,
the relativistic regime. Second, the neutrino masses were assumed to
only depend on the background cosmon $\bar \varphi$ rather than on the
local cosmon value $\varphi$; this is a good approximation as long as
the local cosmon perturbations are sufficiently small, \ie
$m_\nu(\varphi) \approx m_\nu(\bar\varphi)$. Inside neutrino lumps,
this no longer holds.

These issues were addressed by a comprehensive simulation method
specifically designed for GNQ \cite{Ayaita12a}. The latter allows for
relativistic neutrinos whose motion is described by the full equation
of motion, \eq{eom}. The local neutrino mass variations are included
by actually solving the nonlinear field equation for the local cosmon
perturbation $\delta\varphi$ (\cf \sec{ngs}). The backreaction effects
(explained in \sec{structure}) are accounted for by solving the
equations for the homogeneous background simultaneously with the
nonlinear perturbations.

Every neutrino particle $p$ with four-velocity $u^\mu$, proper time
$\tau$ and trajectory $x^\mu_p$ gives rise to an energy-momentum
contribution
\begin{equation}
	T^{\mu\nu}_{(p)}(x) = \frac{1}{\sqrt{-g}} \int \dd \tau\,
	m_\nu(\varphi(x_p))\, u^\mu u^\nu \, \delta^4(x - x_p)
	\label{}
\end{equation}
with the determinant $g$ of the metric and the Dirac delta function.
From this, not only the perturbations of the energy density $\delta
\rho_\nu$ and of the pressure $\delta p_\nu$, the anisotropic shear
${\Sigma^i}_j$, but also the background quantities ${\bar \rho}_\nu =
-{\bar T}^{0}_{(\nu)\,0}$ and ${\bar p}_\nu = {\bar T}^i_{(\nu)\,i}/3$
can be calculated as sums over particle contributions
\cite{Ayaita12a}. These are the actual cosmological averages
\begin{equation}
	{\bar T}^{\mu\nu}_{(\nu)}
	= \frac{\int_{V}^{}\dd^3 x \sqrt{g_{(3)}}\,
	T^{\mu\nu}_{(\nu)}}{\int_{V}^{} \dd^3 x\sqrt{g_{(3)}}},
	\label{}
\end{equation}
that appear in the background equations. Here, $g_{(3)}$ is the
determinant of the spatial metric, and $V$ is the comoving simulation
volume. In this way, the background quantities are directly linked to
the perturbed quantities, thereby including the backreaction effects.

The anisotropic shear ${\Sigma^i}_j$ is no longer negligible once the
neutrinos reach relativistic velocities. Assuming the Newtonian gauge
\begin{equation}
	\dd s^2 = - (1 + 2\Psi) \dd t^2 + a^2 (1 - 2\Phi) \dd \vec x^2,
	\label{}
\end{equation}
it implies a difference $\Psi \not= \Phi$ between the two
gravitational potentials. This is accounted for by solving the
well-known Poisson equation for $\Phi - \Psi$.

The simulation includes also CDM as nonrelativistic particles
accelerated by the Newtonian gravitational potential $\Psi$. In GNQ,
also the neutrino perturbations contribute to $\Psi$ such that
additional forces act on CDM particles potentially increasing their
peculiar velocities \cite{Ayaita09}.

The $N$-body simulation is specified by a number of physical and
numerical parameters \cite{Ayaita12a}. The most important numerical
parameters are the comoving box volume $V = L^3$, the number $N_\nu$
of neutrino and $N_m$ of effective CDM particles, the initial scale
factor $a_i$, and the resolution, \ie the number $N_c$ of cells. A
fixed, equilateral cubic lattice is used. This is sufficient as
cosmon-neutrino structures form on relatively large scales. On this
lattice, the gravitational potentials $\Psi$, $\Phi$, and the cosmon
perturbation $\delta\varphi$ are calculated.

The initial conditions at $a_i$ are taken from linear perturbation
theory \cite{Mota08}. The evolution of CDM particles in the $N$-body
simulation starts even earlier than $a_i$ since nonlinearities in CDM
perturbations occur at much smaller scale factors than in the neutrino
perturbations. The simulation is governed by a global time parameter
for which we use the scale factor $a$. As the dynamical time scale of
the cosmon-neutrino interaction varies with the coupling $\beta$, it
is adequate to let the time steps depend on $\beta$, \ie $\Delta a =
\Delta a(\beta)$.

\subsection{The cosmon field equation}
\label{sec:ngs}

A technical difficulty lies in nonlinearities in the field equation
for cosmon perturbations $\delta\varphi$. Whereas the perturbation
$\delta\varphi$ generally remains rather small, the steepness of the
mass function $m_\nu(\varphi)$ expressed by the large values of
$\beta$ can invalidate the linear approximation
\begin{equation}
	m_\nu(\varphi) \approx m_\nu(\bar \varphi) - \beta(\bar \varphi)\,
	m_\nu(\bar \varphi)\, \delta\varphi.
	\label{}
\end{equation}
This can arise for two reasons. First, if stable cosmon-neutrino lumps
form, the local neutrino mass within lumps effectively freezes while
the mass far outside the lumps continues to grow (\cf
\fig{backreaction}). Second, for very large coupling parameters, \eg
for $\varphi$ close to $\varphi_\text{crit}$ in \eq{divbeta}, the
linear approximation of the mass function can even break down without
a mass-freezing effect.

The nonlinear mass function enters the field equation for
$\delta\varphi$ by virtue of the trace $T_{(\nu)} \propto
m_\nu(\varphi)$. The equation for $\delta\varphi$ is obtained from the
fundamental field equation for $\varphi$, \eq{kg}, which we split into
a homogeneous and a perturbative part. The homogeneous part reads
\begin{equation}
	\ddot{ \bar \varphi } + 3 H \dot{ \bar \varphi } +
	V_{,\varphi}(\bar \varphi) = - \overline{ \beta(\varphi) \,
	T_{(\nu)} }.
	\label{eq:kghom}
\end{equation}
In the perturbative part, we neglect the time derivatives of
$\delta\varphi$ and the nonlinearities in $\delta\varphi$ except for
the coupling parameter and the mass function \cite{Ayaita12a}:
\begin{equation}
	\frac{1}{a^2}\Delta \delta\varphi - V_{,\varphi\varphi}(\bar
	\varphi) + 2 \Psi ( \ddot {\bar \varphi} + 3 H \dot {\bar
	\varphi})
	= \delta\left( \beta(\varphi) T_{(\nu)} \right).
	\label{eq:kgpert}
\end{equation}
Here, the right-hand side is defined as the perturbation
\begin{equation}
	\delta\left( \beta(\varphi) T_{(\nu)} \right)
	= \beta(\varphi) T_{(\nu)} - \overline{ \beta(\varphi) \,
	T_{(\nu)} }
	\label{}
\end{equation}
and can be highly nonlinear in $\delta\varphi$. The solution of
\eq{kgpert} by an iterative Fourier-based method broke down once the
nonlinearities became severe; for the constant coupling model, this
happened at $a \gtrsim 0.5$ \cite{Ayaita12a}. We have implemented a
Newton-Gauß-Seidel (NGS) multigrid relaxation method recently
developed for modified gravity \cite{Puchwein13} to overcome these
difficulties. Thereby, stable solutions of the cosmological evolution
can be obtained.

We write \eq{kgpert} schematically as
\begin{equation}
	\mathcal L[\delta\varphi] \equiv \Delta \delta\varphi -
	F[\delta\varphi] = 0
	\label{}
\end{equation}
with nonlinear functionals $\mathcal L$ and $F$. The NGS solver
applies an iterative prescription which, similarly to Newton's method,
bases upon finding the root of the linearized functional in each
iteration step. However, the linearization is done at every lattice
point $\vec x$ individually; no functional derivative is performed.
The main step of the iteration is thus
\begin{equation}
	\delta\varphi^{(n+1)}(\vec x) = \delta \varphi^{(n)}(\vec x) -
	\frac{\mathcal L[\delta\varphi^{(n)}](\vec x)}{\partial\mathcal
	L[\delta\varphi]/\partial\delta\varphi(\vec x)},
	\label{}
\end{equation}
where the derivative in the denominator is just a usual partial
derivative with respect to the value $\delta\varphi(\vec x)$. The
coupling between neighboring cells is accounted for by the iterative
procedure. We split the derivative as follows:
\begin{equation}
	\frac{\partial\mathcal
	L[\delta\varphi]}{\partial\delta\varphi(\vec x)}
	=
	\frac{\partial\left( \Delta\delta\varphi(\vec
	x)\right)}{\partial\delta\varphi(\vec x)}
	-
	\frac{\partial
	F[\delta\varphi]}{\partial\delta\varphi(\vec x)}
	\label{}.
\end{equation}
Approximating the Laplacian by a seven-point stencil gives us
$-6/\Delta x^2$ for the first term on the right-hand side if $\Delta
x$ is the comoving lattice spacing. In the second term, the crucial
$\delta\varphi$ dependence comes from the product $\beta\,m_\nu$,
\begin{equation}
	\frac{\partial\left[ \beta(\varphi)\, m_\nu(\varphi)
	\right]}{\partial\delta\varphi}
	= \beta_{,\varphi}(\varphi) m_\nu(\varphi) - \beta^2(\varphi)\,
	m_\nu(\varphi).
	\label{}
\end{equation}

For the varying coupling model investigated in this work,
\eq{divbeta}, the coupling $\beta(\varphi)$ and the mass function
$m_\nu(\varphi)$ grow large for $\varphi \to \varphi_\text{crit}$.
When the background cosmon $\bar \varphi$ is very close to the barrier
$\varphi_\text{crit}$, the perturbation $\delta\varphi$ has to be
calculated very accurately. A small numerical error might lead to
exceeding the barrier, $\bar \varphi + \delta\varphi >
\varphi_\text{crit}$, which gives unphysical results. If this is an
issue, a change of variables is appropriate that automatically
enforces the barrier $\varphi < \varphi_\text{crit}$. This is achieved
by solving the field equation for the variable $u(\vec x)$ defined by
\begin{equation}
	\ee^{u(\vec x)} \equiv \varphi_\text{crit} - \varphi(\vec x).
	\label{}
\end{equation}
Regardless of which values $u(\vec x)$ obtains in the NGS solver,
calculating back to $\delta\varphi(\vec x)$ will give a value
respecting the barrier. The resulting term $\Delta e^{u (\varphi(\vec
x))}$ is represented by finite differences as proposed by
Ref.~\cite{Oyaizu08}. The NGS solver uses multigrid acceleration and
the so-called full approximation scheme, which is suited for highly
nonlinear problems. Full details are given in Ref.~\cite{Puchwein13}.


\section{Results}
\label{sec:results}

The generic finding of our simulations is a strong oscillatory
behavior of the neutrino perturbations~-- mildly nonlinear neutrino
overdensities continuously form and dissolve. In contrast to the
stable neutrino lumps in the constant coupling model (\cf
\sec{structure}), these short-lived overdensities never reach high
density contrasts. So, neither do they induce a large gravitational
potential comparable to that of cold dark matter nor do they decouple
from the evolution of the homogeneous background. The expansion
dynamics is only slightly affected. In particular, the effective
average cosmon-neutrino coupling differs only mildly from the
microscopic coupling $\beta$. A standard epoch of accelerated
expansion results from the effective stop of the cosmon evolution.

The numerical method (\cf \sec{method}), however, is not yet
sufficiently fast and robust to explore the parameter space of the
field-dependent coupling model, \eq{divbeta}. A crucial parameter is
the normalization $\bar m$ of the average neutrino mass, defined in
\eq{divmass}. For large $\bar m$, the cosmological evolution becomes
more similar to the constant coupling case. The short-lived
overdensities are more concentrated and massive, and a reliable
numerical treatment of the violent oscillatory behavior in combination
with these concentrated lumps has not yet succeeded. For small $\bar
m$, the neutrinos are lighter and accelerate to highly relativistic
velocities in the process of the formation and dissolution of the
short-lived overdensities. Our method is not yet capable of accurately
resolving the collective motion of neutrinos very close to the speed
of light.

The results presented at this stage are thus obtained for an exemplary
set of parameters. They will be followed by more comprehensive studies
once the numerical methods are sufficiently refined. The neutrino mass
parameter $\bar m$ is chosen as $\bar m = 0.5\times 10^{-3}$~eV
corresponding to a present-day neutrino mass $m_\nu(t_0) \approx
0.2$~eV. In the exponential potential of the cosmon, we choose $\alpha
= 10$. The comoving box of size $V = (600 h^{-1}\text{Mpc})^3$ is
divided into $N_c = 128^3$ cells. The number of effective neutrino and
matter particles is chosen equal to the number of cells, $N_\nu = N_m
= N_c$. The simulation starts for matter at $a_{\text{ini},m} = 0.02$
and adds neutrinos at $a_{\text{ini},\nu} = 0.15$. The initial
perturbations are characterized by a nearly-scale invariant spectrum,
$n_s = 0.96$, with scalar amplitude $A_s = 2.3\times 10^{-9}$ at the
pivot scale $k_\text{pivot} = 0.05\,\text{Mpc}^{-1}$.

\subsection{Cosmic neutrinos}
\label{sec:neutrinos}

\begin{figure}
	\begin{center}
		\includegraphics[width=0.45\textwidth]{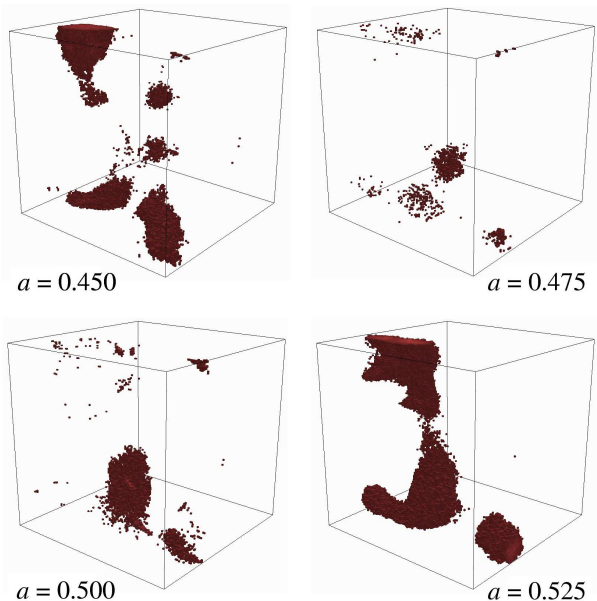}
	\end{center}
	\caption{Forming and dissolving mildly nonlinear neutrino
	overdensities. Simulation cells with a neutrino number density
	contrast $n_\nu/{\bar n}_\nu \geq 5$ are marked red.}
	\label{fig:film}
\end{figure}

One cycle of disappearance and reappearance of mildly nonlinear
neutrino overdensities is shown in \fig{film}.
Large-scale neutrino lumps have formed at $a = 0.45$. At the
intermediate scale factor $a = 0.475$, however, the neutrino
distribution again is almost homogeneous. Shortly afterwards, the
overdensities appear again. Even in their centers, neutrino lumps
hardly reach density contrasts above order $10$. The number of
structures within the whole $600 h^{-1}$Mpc simulation box is very
small. The overdensities thus form on a scale of roughly $\lambda \sim
100h^{-1}$Mpc. This is similar to the constant coupling model in
which, however, the lumps subsequently shrink to the size of several
Megaparsecs and subhalos form \cite{Ayaita12a}. In order to guarantee
that the simulation box is a representative cosmological volume and to
generally avoid box size effects, a larger simulation box would be
desirable. Due to the corresponding loss of resolution or, if more
cells are used, the increased numerical effort, this analysis is
postponed to future work. Our preliminary tests indicate that our
findings are robust with respect to box size.

A period of overdensity formation is initiated by a low neutrino
equation of state $w_\nu \approx 0$. As discussed in the context of
the neutrino equation of motion, \eq{eom}, the bending of neutrino
trajectories and therefore the formation of neutrino overdensities is
most effective in this case. The effect is strengthened when the
cosmon $\bar \varphi$ has come close to the critical value
$\varphi_\text{crit}$ implying large neutrino masses by \eq{divmass}.
Thereafter, the cosmon ``bounces'' against the barrier, and
$\dot{\bar\varphi}$ switches sign becoming negative. The cosmon
acceleration $\propto \beta \dot \varphi - 2H$ in \eq{cosmonacc1}
becomes then positive. Rather than as a damping, it acts as an
accelerant and leads to relativistic neutrino velocities high enough
such that the neutrinos fly out of the lumps. Consequently, a period
of lump formation is followed by a period of lump dissolution.
Subsequently, $\dot{\bar\varphi}$ turns again positive due to the
gradient of the cosmon potential and a new period of lump formation
begins.

\begin{figure}
	\begin{center}
		\psfrag{xlabel}[B][c][.9][0]{scale factor $a$}
		\psfrag{ylabel}[B][c][.9][0]{average neutrino equation of
		state $w_\nu$}
		\includegraphics[width=0.45\textwidth]{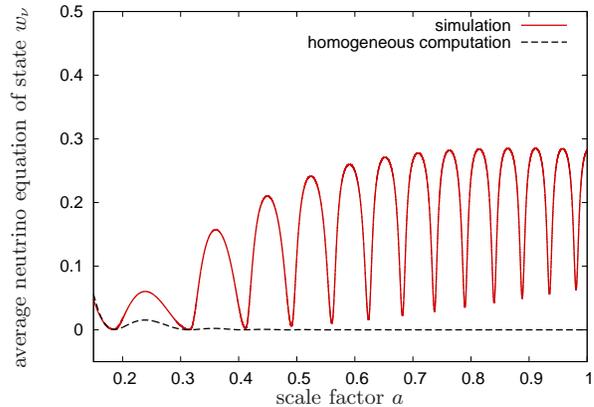}
	\end{center}
	\caption{Evolution of the average equation of state $w_\nu$ of
	neutrinos in the simulation (red solid line) as compared to a
	purely homogeneous computation (black dashed line).}
	\label{fig:wnu}
\end{figure}

These cycles of slow-down and speed-up are visualized in \fig{wnu}.
None of these oscillatory features are visible in a purely homogeneous
calculation, which would predict a neutrino equation of state very
close to zero. It is the proper treatment of nonlinear perturbations
that uncovers why the instability of neutrino perturbations is not
``catastrophic'' \cite{Afshordi05}. The instability, only present for
nonrelativistic neutrinos and leading to the formation of neutrino
lumps, is counteracted by the neutrinos turning relativistic again.
This constitutes a nonlinear stabilization mechanism.

\subsection{Dark Energy}
\label{sec:dark}

\begin{figure}
	\begin{center}
		\psfrag{xlabel}[B][c][.9][0]{scale factor $a$}
		\psfrag{ylabel}[B][c][.9][0]{cosmon equation of state
		$w_\varphi$}
		\includegraphics[width=0.45\textwidth]{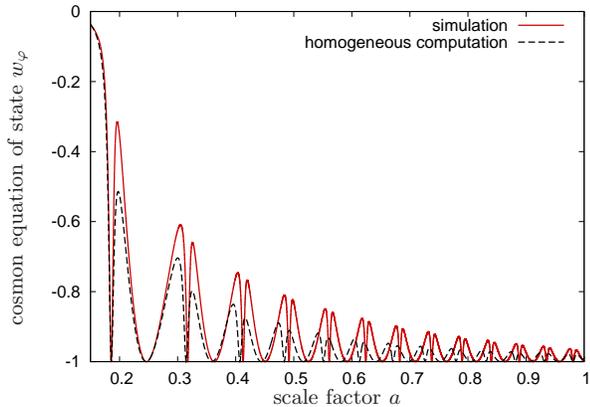}
	\end{center}
	\caption{Evolution of the equation of state $w_\varphi$ of Dark
	Energy in the simulation (red solid line) and in a homogeneous
	computation (black dashed line). The double peak structure
	reflects the oscillation of $\varphi$ in an effective potential.}
	\label{fig:wphi}
\end{figure}

The periods of nonrelativistic neutrino velocities, $w_\nu \approx 0$,
are essential for stopping the evolution of the cosmon and ensuring a
phase of accelerated expansion (\cf \sec{basics}). The periodically
reached relativistic neutrino velocities render the stopping mechanism
slightly less effective as they suppress the coupling term $\propto
T^\lambda_{(\nu)\,\lambda}$ in \eq{kg}. This is visible in the
evolution of the cosmon equation of state $w_\varphi$, \cf \fig{wphi}.
The equation of state $w_\varphi$ approaches the cosmological constant
value $w_\Lambda = -1$ although the full simulation taking the effect
of periodically relativistic neutrino velocities into account
approaches this value a bit more slowly. Albeit clearly visible, this
backreaction effect does not significantly postpone the onset of the
accelerated expansion as seen in \fig{omegas}.

The oscillations in \fig{wphi} show a simple pattern. Repeatedly,
$w_\varphi$ reaches the value $-1$. These are turning points where
$\dot{\bar \varphi}$ switches sign and thus encounters a zero where
$w_\varphi = -1$ holds exactly. Narrow and wide minima alternate. The
narrow minima occur when $\bar \varphi$ bounces against the steep
barrier at $\varphi_\text{crit}$. The wide minima are related to the
other turning point when $\bar \varphi$ has climbed the gently
inclined scalar self-interaction potential $V(\bar\varphi)$. The decay
of the oscillation amplitude for growing $a$ is a consequence of the
damping term $3 H \dot{\bar\varphi}$ in \eq{kghom}.

\begin{figure}
	\begin{center}
		\psfrag{xlabel}[B][c][.9][0]{scale factor $a$}
		\psfrag{ylabel}[B][c][.9][0]{coupling parameter $\beta$}
		\includegraphics[width=0.45\textwidth]{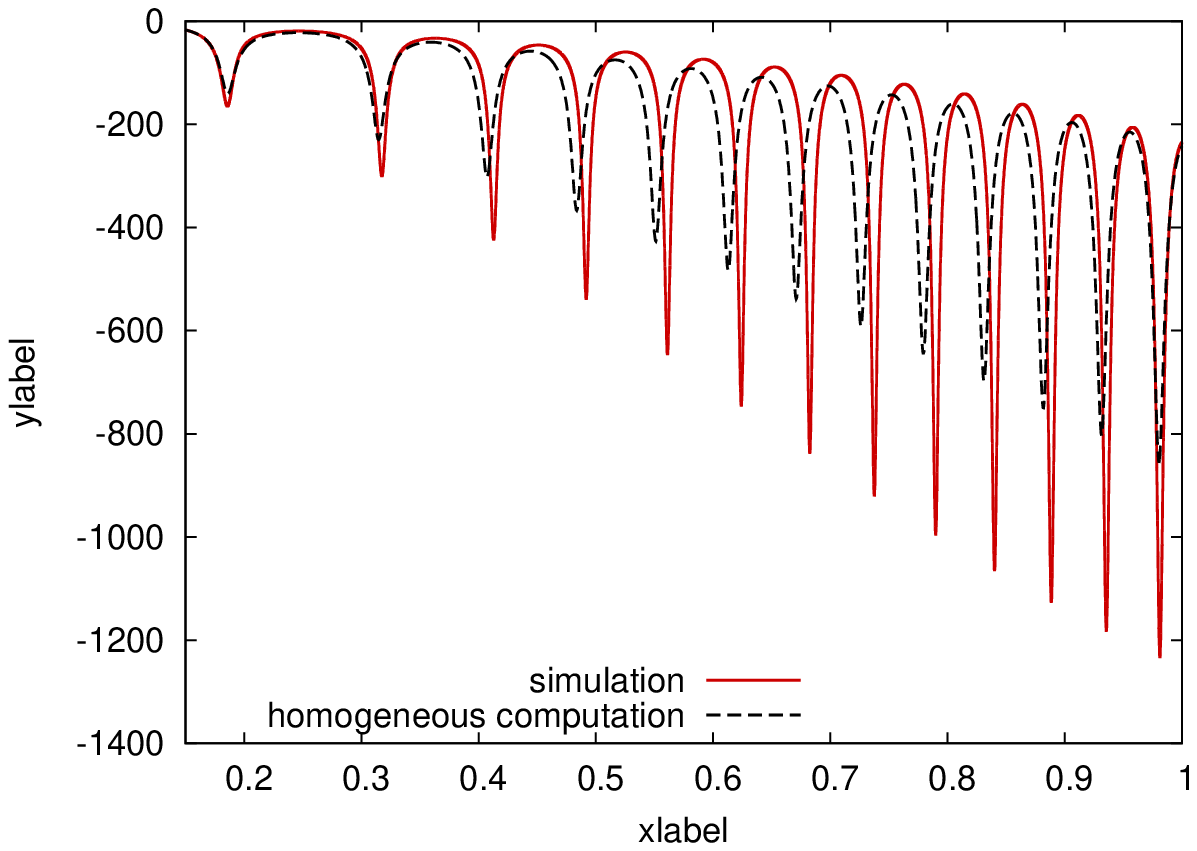}
		\psfrag{xlabel}[B][c][.9][0]{scale factor $a$}
		\psfrag{ylabel}[B][c][.9][0]{average neutrino mass ${\bar
		m}_\nu$ [eV]}
		\includegraphics[width=0.45\textwidth]{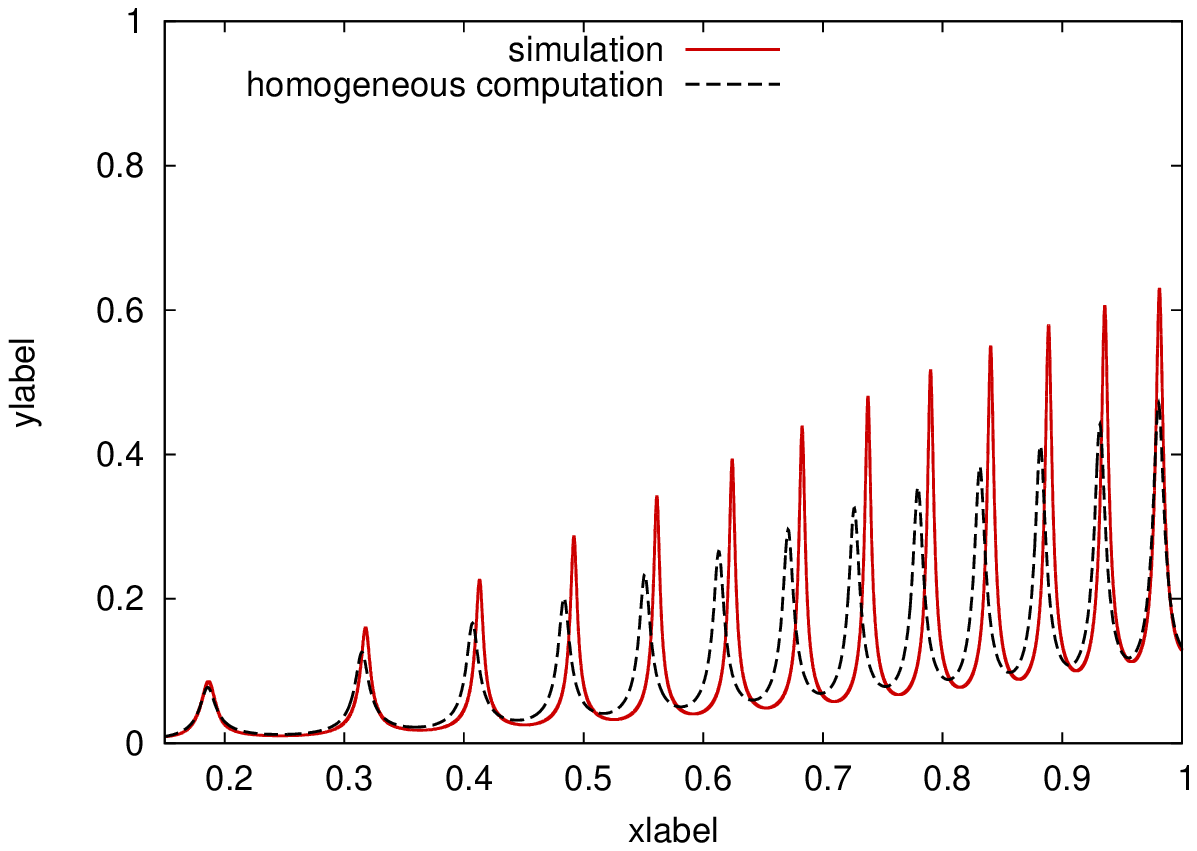}
	\end{center}
	\caption{Evolution of the coupling parameter $\beta(\varphi)$
	and the average neutrino mass $\overline{m_\nu(\varphi)}$
	in the simulation (red solid line) as compared to a purely
	homogeneous computation (black dashed line).
	The nonlinear effects
	(backreaction) enhance the peaks and lead to a small shift.}
	\label{fig:beta}
	\label{fig:mnu}
\end{figure}

The evolution of the cosmon $\bar \varphi$ is reflected in the
evolution of the coupling parameter $\beta(\bar \varphi)$ and the
average neutrino mass $m_\nu(\bar \varphi)$, \cf \fig{beta}.
They are both proportional to the inverse of $\varphi_\text{crit} -
\bar \varphi$ (which is the distance to the barrier),
Eqs.~(\ref{eq:divbeta}) and (\ref{eq:divmass}), and reach extrema at
the turning points of $\bar \varphi$. As the cosmon perturbations
$\delta\varphi$ remain small, it is justified to assume $\overline{
\beta(\varphi)} \approx \beta(\bar\varphi)$ and $\overline{
m_\nu(\varphi) } \approx m_\nu(\bar \varphi)$ for the averages; this
is different from the constant coupling case, \cf \sec{structure}.
Figure~\ref{fig:beta} tells us that the backreaction effect is most
pronounced at the cusps of the plots. The coupling reaches values
$\beta \approx -1.2 \times 10^3$, and the highest average neutrino
mass is $m_\nu \approx 0.6$~eV. At the opposite point of the
oscillation, the coupling parameter is around $\beta \approx -2\times
10^2$, and the mass is at $m_\nu \approx 0.15$~eV. As the precise
oscillatory pattern will sensitively depend on the chosen model
parameters, we conclude that the varying coupling model will not
predict a precise value for the present-day neutrino mass but rather a
range.

\subsection{Neutrino lump gravitational potential}
\label{sec:potential}

\begin{figure}
	\begin{center}
		\psfrag{xlabel}[B][c][.9][0]{scale factor $a$}
		\psfrag{ylabel}[B][c][.9][0]{relative potential
		$\Phi_\nu(k)/\Phi(k)$}
		\includegraphics[width=0.45\textwidth]{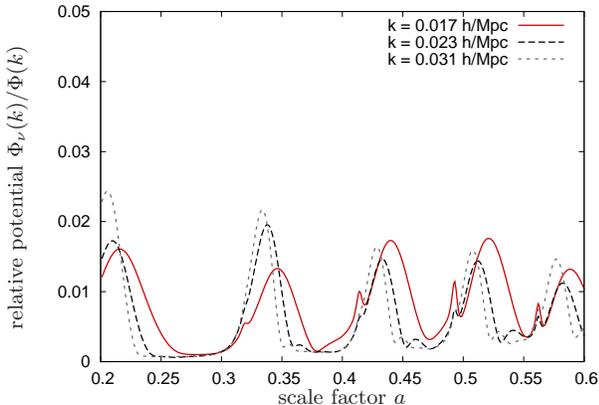}
	\end{center}
	\caption{Comparison between the neutrino-induced gravitational
	potential $\Phi_\nu$ and the total gravitational potential $\Phi$
	during the first oscillations at three large comoving scales.}
	\label{fig:potentials}
\end{figure}

The only way for cosmological observations to detect the large
neutrino overdensities is via the effects of their gravitational
potentials. These gravitational potentials have an impact on the
evolution of CDM perturbations, in particular on the peculiar velocity
field \cite{Ayaita09}. More importantly, they can leave direct
observational traces on the cosmic microwave background via the ISW
effect. The quantitative results on these gravitational potentials
will thus ultimately decide whether the GNQ model will prove viable in
the light of various observational constraints. Although answering
this question is beyond the scope of this work as it requires an
exploration of the model's parameter space, we show the results
obtained for the exemplary set of parameters employed here in
\fig{potentials}.
The neutrino-induced gravitational potential $\Phi_\nu$ is
subdominant, by two orders of magnitude, compared to the CDM potential
$\Phi_m$. More precisely, the figure shows the dimensionless spectra
$\Phi_\nu(k)$, $\Phi_m(k)$ (\cf also Ref.~\cite{Ayaita12a}).

Between the different comoving scales, we observe a phase shift.
During the dissolution process, small scales are washed out more
rapidly than large scales. Inversely, during the formation process,
small-scale perturbations build up faster.

\section{Physical picture}
\label{sec:interpretation}

\subsection{Effective cosmon dynamics}
\label{sec:effcosmon}

Our aim is an analytic understanding of the time evolution of the
average cosmon field $\bar \varphi$ in the presence of inhomogeneous
and possibly rapidly moving neutrinos. For this purpose, we employ an
effective potential $V_\text{eff}(\bar \varphi)$ which depends, in
addition to $\bar\varphi$, also on a characteristic neutrino momentum
$p$ and the average neutrino density ${\bar n}_\nu$. Both, $p$ and
${\bar n}_\nu$ may depend on the scale factor $a$ or other
cosmological quantities, but are assumed to show no explicit
dependence on $\bar \varphi$. The time evolution of $\bar \varphi$
will then be governed by the equation of motion
\begin{equation}
	\ddot{ \bar \varphi } + 3 H \dot{ \bar \varphi } +
	V_{\text{eff}\, ,\bar\varphi}(\bar \varphi) = 0.
	\label{eq:kgeff}
\end{equation}
The derivative $V_{\text{eff}\, ,\bar\varphi}$ is composed of the
self-interaction part $V_{,\bar\varphi}$ and a contribution from the
cosmon-neutrino interaction given by the right-hand side of
\eq{kghom},
\begin{equation}
	V_{\text{eff}\, ,\bar\varphi}(\bar\varphi)
	= V_{,\bar\varphi}(\bar\varphi) + \overline{ \beta T_{(\nu)} }.
	\label{}
\end{equation}
For an estimate of the coupling term, we assume that it can be written
in the form
\begin{equation}
	\overline{ \beta T_{(\nu)} }
	= \frac{\partial {\bar m}_\nu(\bar \varphi)}{\partial \bar
	\varphi}\, \frac{ {\bar n}_\nu}{\bar \gamma},
	\label{eq:a1}
\end{equation}
with ${\bar m}_\nu(\bar \varphi)$ the average neutrino mass and ${\bar
n}_\nu(a) \propto a^{-3}$ the (known) average neutrino number density.
The exact formula would be a sum over individual particle
contributions, where the right-hand side for each particle is just as
in \eq{a1} if we replace $\bar \gamma$ by the usual Lorentz factor
$\gamma$ (\cf Ref.~\cite{Ayaita12a}). The effective Lorentz factor
$\bar \gamma$ is assumed to depend on $\bar \varphi$ only through
${\bar m}_\nu(\bar \varphi)$. Then, dimensional analysis implies that
$\bar \gamma$ is a function of the combination $p^2 / {\bar
m}_\nu^2(\bar \varphi)$, where $\vec p$ is some appropriate
characteristic momentum for the neutrinos and $p = \sqrt{\vec p^2}$.
In principle, the difference between the average of $\partial m_\nu /
\partial \varphi$ and $\partial {\bar m}_\nu / \partial \bar \varphi$
is included in the factor $\bar \gamma$. For the present scenario, our
numerical simulations show that these two quantities are approximately
equal since the neutrino density perturbation sourcing the cosmon
perturbation $\delta\varphi$ never reaches large values during the
cosmic evolution.

The effective potential can now be defined as
\begin{equation}
	V_\text{eff}(\bar \varphi) = V(\bar \varphi)
	+ {\bar n}_\nu \, {\bar m}_\nu(\bar \varphi)\, \hat \gamma,
	\label{eq:a2}
\end{equation}
with $\hat \gamma$ related to $\bar \gamma$ by
\begin{equation}
	\hat \gamma +
	\frac{\partial \hat \gamma}{\partial \ln {\bar m}_\nu}
	= \frac{1}{\bar \gamma}.
	\label{eq:a3}
\end{equation}
Employing that $\hat \gamma$ is a dimensionless function of $p^2 /
{\bar m}_\nu^2$, \eq{a3} follows directly from \eq{kgeff} and the
definition of $\bar \gamma$ by \eq{a1}. For the case of a free
particle with
\begin{equation}
	{\bar \gamma}^2 = 1 + \frac{p^2}{ {\bar m}_\nu^2(\bar
	\varphi)},
	\label{eq:a4}
\end{equation}
one obtains $\hat \gamma = \bar \gamma$. For more general momentum
distributions of neutrinos, the functions $\bar \gamma(p^2 / {\bar
m}_\nu^2)$ and $\hat \gamma (p^2 / {\bar m}_\nu^2)$ may be somewhat
more complicated, but the qualitative relation remains similar.

We next need to understand the time evolution of the characteristic
neutrino momentum $p$. We distinguish for each oscillation period two
stages. The first stage is characterized by the importance of
inhomogeneities in the cosmon field, occurring when $\bar\varphi$ is
close to the critical value $\varphi_\text{crit}$. There, the neutrino
mass is close to its maximum, and $p^2 \ll {\bar m}_\nu^2 (\bar
\varphi)$ such that $\bar \gamma \approx \hat \gamma \approx 1$. The
spatial cosmon gradients in the neutrino equation of motion, \eq{eom},
lead to the growth of $p^2$ and to the formation of neutrino
overdensities. We identify a second stage when $\bar\varphi$ is no
longer close to $\varphi_\text{crit}$. Here, inhomogeneities in the
cosmon field are no longer decisive, and the overall dynamics is
dominated by cosmon acceleration. We will argue that, for this second
stage, $p^2$ is (almost) conserved. The effective potential
$V_\text{eff}(\bar\varphi)$ then only depends on the value of $p^2$
that has been reached during the first stage.

For a single particle in a homogeneous background, the combination
$\vec p / a$ is conserved due to translation symmetry. In the absence
of gravity (for constant $a$), this follows directly from the neutrino
equation of motion, \eq{eom}, in the case where spatial gradients of
$\varphi$ can be neglected,
\begin{equation}
	\frac{\dd u^k}{\dd t} = \beta(\bar \varphi) \,
	\dot{\bar\varphi}\, u^k.
	\label{}
\end{equation}
This equation of motion conserves the relativistic momentum $\vec p =
m_\nu(\bar \varphi)\, \vec u$. Beyond small effects from the
expanding scale factor, any additional change of $p$ has to arise from
inhomogeneities. These are small during the second stage.

\begin{figure}[h]
	\begin{center}
		\psfrag{xlabel}[B][c][.9][0]{cosmon value $\bar \varphi -
		\varphi_\text{crit}$}
		\psfrag{ylabel}[B][c][.9][0]{effective potential
		$V_\text{eff}(\bar\varphi)/V(0)$}
		\includegraphics[width=0.45\textwidth]{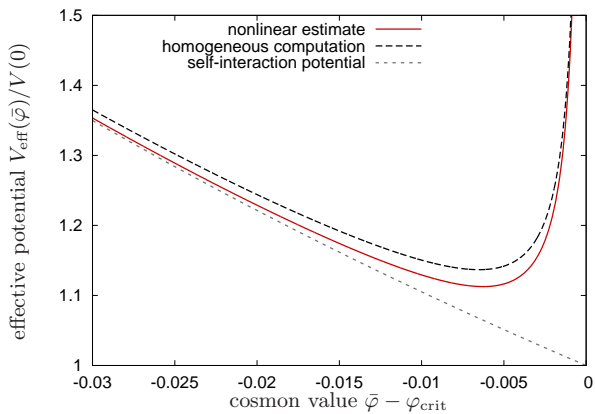}
	\end{center}
	\caption{Effective cosmon potential $V_\text{eff}(\bar\varphi)$
	for the nonlinear (red solid) and the homogeneous case (black
	dashed). For comparison, we also show the self-interaction
	potential $V(\bar \varphi)$ (gray dotted) without the effects of
	the cosmon-neutrino coupling. Units of $V$ are set by the
	normalization factor $V(0) = 1.06\times
	10^{-7}\text{Mpc}^{-2}$.}
	\label{fig:veff}
\end{figure}

We plot, in \fig{veff}, the effective potential given by \eq{a4},
taking the parameters from the simulation around the oscillation at $a
\approx 0.5$.
The homogeneous computation that we show as a comparison employs $p =
0$. The cosmon $\bar \varphi$ oscillates around the minimum of the
effective potential according to the effective equation of motion,
\eq{kgeff}. The asymmetry of the effective potential explains the
double peak structure discussed in \sec{dark}.

The identification of the characteristic neutrino momentum $p$ as the
decisive parameter for the cosmon dynamics opens the door to effective
descriptions no longer relying on a full cosmological simulation. For
example, the momentum build-up during overdensity formation might be
estimated in a suitable spherical collapse approach
\cite{Wintergerst10b} or even with an adaption of linear perturbation
theory. A detailed investigation of these routes is beyond the scope
of this note and left for future work.

\subsection{Effective neutrino dynamics}
\label{sec:effneutrino}

Within the structure formation cycle, we consider first the period of
approximate homogeneity for which the cosmon acceleration has
violently disrupted the previously formed overdensities. During this
period, the effects of the mutual attraction of the neutrinos are
suppressed due to their relativistic velocities. The decisive
conserved quantity is the relativistic momentum $p$ whose value is
determined by the preceding period of overdensity formation. Of
course, $p$ shrinks due to the ordinary Hubble damping; but this
effect is small because it is linked to a much larger time scale,
$1/H$, as compared to the dynamic timescale of the cosmon-neutrino
coupling, which is $1/|\beta \dot{\bar \varphi}|$. During this
approximately homogeneous phase, the neutrinos influence the cosmon
evolution via the effective potential $V_\text{eff}(\bar\varphi;p)$
according to \eq{kgeff}. For large enough $\varphi_\text{crit} - \bar
\varphi$, there will be a turnaround with $\bar\varphi$ increasing
subsequently until a new phase of lump formation sets in.

We next discuss the phase of lump formation. During this phase, the
influence of neutrino inhomogeneities on the effective potential
$V_\text{eff}(\bar\varphi; p)$ is small, and we may use the
homogeneous computation ($p = 0$). Indeed, when $\bar\varphi$ comes
close to $\varphi_\text{crit}$ (where overdensities form), the
neutrinos become nonrelativistic due to their rapidly growing masses,
and the homogeneous computation of $V_\text{eff}$ is fairly accurate.
In principle, the validity of the homogeneous computation could be
spoiled by the type of backreaction effects encountered in the
constant coupling model, \cf \sec{structure}, where the local cosmon
value effectively freezes and no longer follows the homogeneous
component. We do not observe these effects, here, since the neutrino
overdensities never become large.

Taking things together, we end with a rather simple qualitative
understanding of the role of nonlinearities. The evolution of the
cosmon average field $\bar \varphi$ is rather independent of the
details of the lump formation process. We only need to understand the
small increase of the characteristic neutrino momentum $p$ during each
phase of lump formation. On the other hand, for the stages of lump
formation, the cosmon field dynamics can be approximated by
neglecting the backreaction effects (\eg $p = 0$ in $V_\text{eff}$).

The details of the stages of lump dissolution are not crucial because
the overall picture there is just given by the conservation of the
momentum distribution $f(\vec p)$. The latter provides an explanation
for some of the observations made in \sec{neutrinos}. Not only did the
lumps periodically appear and disappear~-- they occurred roughly in a
similar shape as in the preceding period. Furthermore, the periodic
minima in the neutrino equation of state $w_\nu$, \cf \fig{wnu}, reach
higher values every time rather than always shrinking to $w_\nu = 0$.
This cannot be explained by the spatial distribution of the neutrino
density which is, to a very good approximation, homogeneous after each
dissolution phase.  It is the conservation of the neutrino momentum
distribution $f(\vec p)$ during the dissolution process that tells us
that the overdensity formation process does not start from a clean
state. When the overdensities are just to form again, the neutrino
velocities are already pointing to the --~previous and next~--
overdensities' centers. The momentum build-up during overdensity
formation then adds up with the preceding momentum. In each iteration,
the momentum thus takes on larger values, and the equation of state
after overdensity formation has increased compared to the last
iteration.

Rather than describing the process of overdensity formation in an
adapted linear perturbation theory or spherical collapse scheme, we
content ourselves here with a brief qualitative discussion explaining
the main effects. This will make plausible our finding that the
neutrino-induced gravitational potentials remain small compared to
those of CDM (\cf \fig{potentials}). A refined analysis will be the
subject of future work. The overdensities form when the cosmon comes
close to $\varphi_\text{crit}$, the barrier in the effective potential
$V_\text{eff}$. The spatial cosmon gradients become important compared
to the time derivative. The coupling parameter $\beta$ reaches order
$-10^3$, \cf \fig{beta}.

Although the resulting forces on the neutrinos are $2\beta^2 \sim
10^{6}$ times stronger than gravity, several factors hinder the
formation of highly concentrated lumps.  First, the period of time
during which the cosmon is close to $\varphi_\text{crit}$ and the
neutrinos are nonrelativistic is limited to roughly $\Delta a \sim
10^{-2}$, \cf Figs.~\ref{fig:wnu} and \ref{fig:wphi}. The
nonrelativistic neutrinos are not fast enough to form overdensities
beyond roughly $\delta_\nu \sim 10^{-2}$ as seen in the simulation.
After the cosmon has bounced against the barrier and
$\dot{\bar\varphi}$ is negative, the cosmon acceleration increases all
neutrino velocities along their respective directions of motion. At
first, the overdensities continue to grow as the neutrino velocities
were on average, during the nonrelativistic period, targeting towards
the centers of the forming overdensities. This explains why the maxima
of the neutrino-induced gravitational potential occur at later times
than the bouncing of $\bar\varphi$ against the barrier, \cf
\fig{potentials}. However, during the cosmon acceleration, the
neutrino masses rapidly decrease. Consequently, although the number
overdensities grow, the effect on the gravitational potential is only
moderate. At its maxima, the large-scale gravitational potential
induced by neutrinos is only at the percent level compared to the CDM
potential.

For small scales, the neutrinos form even less overdensities compared
to $\delta_m$. The relative importance of the neutrino gravitational
potential therefore decreases towards smaller length scales.


\section{Conclusion}
\label{sec:conclusion}

The Growing Neutrino Quintessence model with a field-dependent
coupling $\beta(\varphi)$ shows violent nonlinear dynamics of the
coupled cosmon-neutrino fluid, and yet an overall phenomenology
similar to the standard $\Lambda$CDM picture. The accelerated
expansion is almost the same as for $\Lambda$CDM (\cf \fig{omegas}),
while large-scale neutrino overdensities remain small enough so that
their induced gravitational potentials are subdominant to those of
cold dark matter. At the fundamental level, however, the model is not
anywhere near $\Lambda$CDM. Rather than being a parameter, the present
Dark Energy density results from the stop of a scaling solution by a
cosmological trigger event, namely neutrinos becoming nonrelativistic.
In the process, the coupling parameter between the neutrinos and Dark
Energy dynamically reaches order $\beta(\varphi) \sim -10^3$ (\cf
\fig{beta}), inducing an attraction between neutrinos $10^6$ times
stronger than gravity. This may serve as an example that a standard
overall phenomenology still leaves room for new physics, without
unnaturally small parameters.

The average neutrino mass $m_\nu$ is small in the early universe. For
most of the cosmological evolution, the Dark Energy scalar field rolls
down steadily its potential towards larger values, and the average
neutrino mass grows with time. In the recent epoch, however, the
cosmon field value oscillates and so do the neutrino masses. For the
investigated parameters, $m_\nu$ oscillates between about $0.15$~eV
and $0.6$~eV (\cf \fig{mnu}). Nonrelativistic neutrinos experience an
attractive force due to the coupling $\beta(\varphi)$ substantially
stronger than gravity. Furthermore, Hubble damping is replaced by
cosmon acceleration.

The violent nonlinear behavior of neutrino perturbations manifests
itself in the repeated rapid formation and dissolution of large-scale
overdensities (\cf \fig{film}). Rather than becoming nonrelativistic
once and for all, the neutrinos are accelerated to relativistic
velocities periodically (\cf \fig{wnu}). This effectively stabilizes
the evolution of perturbations, and the ``catastrophic'' instability
first discussed in the context of MaVaNs is avoided \cite{Afshordi05}.
The (short-lived) overdensities induce oscillating gravitational
potentials, whose relative strength compared to those of cold dark
matter remains at the percent level (\cf \fig{potentials}).

By virtue of the coupling between Dark Energy and the neutrinos, the
nonlinear dynamics of neutrino perturbations exert a backreaction
effect on the evolution of Dark Energy at the background level.
Relativistic neutrino velocities reduce the strength of the effect of
the coupling and thereby weaken the Dark Energy stopping mechanism.
Although the backreaction is clearly visible quantitatively for the
equations of state of individual components, it does not alter the
qualitative picture with a usual crossover to the accelerated
expansion epoch (\cf \fig{wphi}). For our parameter set, the
backreaction effect becomes negligible for the evolution of the
overall energy fraction for the coupled cosmon-neutrino fluid which
constitutes Dark Energy, see \fig{omegas}. This finding is in contrast
to the constant coupling model in which stable neutrino lumps form and
effectively decouple from the homogeneous component. There, a much
stronger backreaction effect substantially postpones the onset of the
accelerated expansion \cite{Ayaita12a}.

We have obtained our numerical results from an $N$-body based
simulation technique \cite{Ayaita12a}, specifically developed for the
Growing Neutrino Quintessence model, together with a
Newton-Gauß-Seidel solver for the local Dark Energy perturbations
\cite{Puchwein13}. Our method (described in \sec{method}) has allowed
to show, for the first time, the nonlinear evolution of the model
until redshift zero. Earlier attempts had to stop at $z \approx 1$ and
were restricted to the technically simpler constant coupling model
\cite{Baldi11a, Ayaita12a}. Still, the very strong coupling parameters
and the violent perturbation evolution have so far prevented a scan of
the model's parameter space for the field-dependent coupling model.
This, however, would be a decisive step towards a confrontation of the
model with observational constraints. Further efforts are required to
render the numerical method faster and more robust.

A complementary road consists in a semi-analytical approach allowing
for a simplified yet reliable description of the cosmological
dynamics. In the constant coupling model, this had inspired the
physical picture of a cosmon-neutrino lump fluid \cite{Ayaita12b}. We
have laid the ground here for such an effective description of the
field-dependent coupling model, whose cornerstones we have explained
in \sec{interpretation}. In periods during which the cosmon evolves
rapidly, the neutrino momenta are approximately conserved. This
conservation has enabled us to define an effective self-interaction
potential for the scalar field $V_\text{eff}(\varphi)$ (\cf
\fig{veff}) that fully describes the evolution of the homogeneous Dark
Energy. For the neutrino component, our findings motivate an adapted
spherical collapse approach that would allow to estimate the momentum
build-up during the overdensity formation process. Such a
semi-analytical approach will be shaped along with the continuing work
on the numerical simulation method.

Despite the important steps still to be done, the overall picture for
the confrontation of Growing Neutrino Quintessence with observations
already takes a clear shape. For the varying coupling $\beta(\varphi)$
and the parameter set chosen for the present paper, the background
evolution is essentially indistinguishable from the $\Lambda$CDM
model. (The small fraction of Early Dark Energy can be further reduced
by a larger value of the parameter $\alpha$.) Also the gravitational
potential induced by the neutrino lumps seems too small for an easy
observational detection. Such models appear to be compatible with
present observations to the same degree as $\Lambda$CDM. On the other
side, the models with constant coupling $\beta$ may allow for
parameters such that the present Dark Energy can be adjusted to the
observed value. In this case, we expect much stronger neutrino-induced
gravitational potentials, observable by the ISW effect or other tests.
It is obvious that, for part of the parameter space, Growing Neutrino
Quintessence deviates substantially from observation and the
$\Lambda$CDM model. Large parameter regions lie between the two
extremes. They will allow for clear signals for future observations
without being inconsistent with present observations. The cosmic
neutrino background may finally become observable.


\begin{acknowledgments}
	The authors are thankful to Valeria Pettorino and Santiago Casas
	for numerous inspiring discussions and ideas. They would also like
	to thank Luca Amendola, Volker Springel, and Maik Weber for
	valuable suggestions, and David Mota for providing his numerical
	implementation of linear perturbation theory in GNQ. FF
	acknowledges support from the IMPRS-PTFS and the DFG through the
	TRR33 project ``The Dark Universe''. MB is supported by the Marie
	Curie Intra European Fellowship ``SIDUN'' within the 7th Framework
	Programme of the European Commission. EP acknowledges support by
	the DFG through TRR33 and by the ERC grant ``The Emergence of
	Structure during the epoch of Reionization''.
\end{acknowledgments}


\bibliography{varbeta}

\begin{thebibliography}{38}%
\makeatletter
\providecommand \@ifxundefined [1]{%
 \@ifx{#1\undefined}
}%
\providecommand \@ifnum [1]{%
 \ifnum #1\expandafter \@firstoftwo
 \else \expandafter \@secondoftwo
 \fi
}%
\providecommand \@ifx [1]{%
 \ifx #1\expandafter \@firstoftwo
 \else \expandafter \@secondoftwo
 \fi
}%
\providecommand \natexlab [1]{#1}%
\providecommand \enquote  [1]{``#1''}%
\providecommand \bibnamefont  [1]{#1}%
\providecommand \bibfnamefont [1]{#1}%
\providecommand \citenamefont [1]{#1}%
\providecommand \href@noop [0]{\@secondoftwo}%
\providecommand \href [0]{\begingroup \@sanitize@url \@href}%
\providecommand \@href[1]{\@@startlink{#1}\@@href}%
\providecommand \@@href[1]{\endgroup#1\@@endlink}%
\providecommand \@sanitize@url [0]{\catcode `\\12\catcode `\$12\catcode
  `\&12\catcode `\#12\catcode `\^12\catcode `\_12\catcode `\%12\relax}%
\providecommand \@@startlink[1]{}%
\providecommand \@@endlink[0]{}%
\providecommand \url  [0]{\begingroup\@sanitize@url \@url }%
\providecommand \@url [1]{\endgroup\@href {#1}{\urlprefix }}%
\providecommand \urlprefix  [0]{URL }%
\providecommand \Eprint [0]{\href }%
\providecommand \doibase [0]{http://dx.doi.org/}%
\providecommand \selectlanguage [0]{\@gobble}%
\providecommand \bibinfo  [0]{\@secondoftwo}%
\providecommand \bibfield  [0]{\@secondoftwo}%
\providecommand \translation [1]{[#1]}%
\providecommand \BibitemOpen [0]{}%
\providecommand \bibitemStop [0]{}%
\providecommand \bibitemNoStop [0]{.\EOS\space}%
\providecommand \EOS [0]{\spacefactor3000\relax}%
\providecommand \BibitemShut  [1]{\csname bibitem#1\endcsname}%
\let\auto@bib@innerbib\@empty
\bibitem [{\citenamefont {Perlmutter}\ \emph {et~al.}(1999)\citenamefont
  {Perlmutter}, \citenamefont {Aldering}, \citenamefont {Goldhaber},
  \citenamefont {Knop}, \citenamefont {Nugent} \emph {et~al.}}]{Perlmutter99}%
  \BibitemOpen
  \bibfield  {author} {\bibinfo {author} {\bibfnamefont {S.}~\bibnamefont
  {Perlmutter}}, \bibinfo {author} {\bibfnamefont {G.}~\bibnamefont
  {Aldering}}, \bibinfo {author} {\bibfnamefont {G.}~\bibnamefont {Goldhaber}},
  \bibinfo {author} {\bibfnamefont {R.}~\bibnamefont {Knop}}, \bibinfo {author}
  {\bibfnamefont {P.}~\bibnamefont {Nugent}},  \emph {et~al.} (\bibinfo
  {collaboration} {Supernova Cosmology Project}),\ }\href {\doibase
  10.1086/307221} {\bibfield  {journal} {\bibinfo  {journal} {Astrophys. J.}\
  }\textbf {\bibinfo {volume} {517}},\ \bibinfo {pages} {565} (\bibinfo {year}
  {1999})},\ \Eprint {http://arxiv.org/abs/astro-ph/9812133}
  {arXiv:astro-ph/9812133} \BibitemShut {NoStop}%
\bibitem [{\citenamefont {Riess}\ \emph {et~al.}(1998)\citenamefont {Riess},
  \citenamefont {Filippenko}, \citenamefont {Challis}, \citenamefont
  {Clocchiatti}, \citenamefont {Diercks} \emph {et~al.}}]{Riess98}%
  \BibitemOpen
  \bibfield  {author} {\bibinfo {author} {\bibfnamefont {A.~G.}\ \bibnamefont
  {Riess}}, \bibinfo {author} {\bibfnamefont {A.~V.}\ \bibnamefont
  {Filippenko}}, \bibinfo {author} {\bibfnamefont {P.}~\bibnamefont {Challis}},
  \bibinfo {author} {\bibfnamefont {A.}~\bibnamefont {Clocchiatti}}, \bibinfo
  {author} {\bibfnamefont {A.}~\bibnamefont {Diercks}},  \emph {et~al.}
  (\bibinfo {collaboration} {Supernova Search Team}),\ }\href {\doibase
  10.1086/300499} {\bibfield  {journal} {\bibinfo  {journal} {Astron. J.}\
  }\textbf {\bibinfo {volume} {116}},\ \bibinfo {pages} {1009} (\bibinfo {year}
  {1998})},\ \Eprint {http://arxiv.org/abs/astro-ph/9805201}
  {arXiv:astro-ph/9805201} \BibitemShut {NoStop}%
\bibitem [{\citenamefont {Bartelmann}(2010)}]{Bartelmann10}%
  \BibitemOpen
  \bibfield  {author} {\bibinfo {author} {\bibfnamefont {M.}~\bibnamefont
  {Bartelmann}},\ }\href {\doibase 10.1103/RevModPhys.82.331} {\bibfield
  {journal} {\bibinfo  {journal} {Rev.Mod.Phys.}\ }\textbf {\bibinfo {volume}
  {82}},\ \bibinfo {pages} {331} (\bibinfo {year} {2010})},\ \Eprint
  {http://arxiv.org/abs/0906.5036} {arXiv:0906.5036 [astro-ph.CO]} \BibitemShut
  {NoStop}%
\bibitem [{\citenamefont {Copeland}\ \emph {et~al.}(2006)\citenamefont
  {Copeland}, \citenamefont {Sami},\ and\ \citenamefont
  {Tsujikawa}}]{Copeland06}%
  \BibitemOpen
  \bibfield  {author} {\bibinfo {author} {\bibfnamefont {E.~J.}\ \bibnamefont
  {Copeland}}, \bibinfo {author} {\bibfnamefont {M.}~\bibnamefont {Sami}}, \
  and\ \bibinfo {author} {\bibfnamefont {S.}~\bibnamefont {Tsujikawa}},\ }\href
  {\doibase 10.1142/S021827180600942X} {\bibfield  {journal} {\bibinfo
  {journal} {Int.J.Mod.Phys.}\ }\textbf {\bibinfo {volume} {D15}},\ \bibinfo
  {pages} {1753} (\bibinfo {year} {2006})},\ \Eprint
  {http://arxiv.org/abs/hep-th/0603057} {arXiv:hep-th/0603057 [hep-th]}
  \BibitemShut {NoStop}%
\bibitem [{\citenamefont {Amendola}\ \emph {et~al.}(2008)\citenamefont
  {Amendola}, \citenamefont {Baldi},\ and\ \citenamefont
  {Wetterich}}]{Amendola08}%
  \BibitemOpen
  \bibfield  {author} {\bibinfo {author} {\bibfnamefont {L.}~\bibnamefont
  {Amendola}}, \bibinfo {author} {\bibfnamefont {M.}~\bibnamefont {Baldi}}, \
  and\ \bibinfo {author} {\bibfnamefont {C.}~\bibnamefont {Wetterich}},\ }\href
  {\doibase 10.1103/PhysRevD.78.023015} {\bibfield  {journal} {\bibinfo
  {journal} {Phys.Rev.}\ }\textbf {\bibinfo {volume} {D78}},\ \bibinfo {pages}
  {023015} (\bibinfo {year} {2008})},\ \Eprint {http://arxiv.org/abs/0706.3064}
  {arXiv:0706.3064 [astro-ph]} \BibitemShut {NoStop}%
\bibitem [{\citenamefont {Wetterich}(2007)}]{Wetterich07}%
  \BibitemOpen
  \bibfield  {author} {\bibinfo {author} {\bibfnamefont {C.}~\bibnamefont
  {Wetterich}},\ }\href {\doibase 10.1016/j.physletb.2007.08.060} {\bibfield
  {journal} {\bibinfo  {journal} {Phys.Lett.}\ }\textbf {\bibinfo {volume}
  {B655}},\ \bibinfo {pages} {201} (\bibinfo {year} {2007})},\ \Eprint
  {http://arxiv.org/abs/0706.4427} {arXiv:0706.4427 [hep-ph]} \BibitemShut
  {NoStop}%
\bibitem [{\citenamefont {Weinberg}(1989)}]{Weinberg89}%
  \BibitemOpen
  \bibfield  {author} {\bibinfo {author} {\bibfnamefont {S.}~\bibnamefont
  {Weinberg}},\ }\href {\doibase 10.1103/RevModPhys.61.1} {\bibfield  {journal}
  {\bibinfo  {journal} {Rev. Mod. Phys.}\ }\textbf {\bibinfo {volume} {61}},\
  \bibinfo {pages} {1} (\bibinfo {year} {1989})}\BibitemShut {NoStop}%
\bibitem [{\citenamefont {Carroll}(2001)}]{Carroll01}%
  \BibitemOpen
  \bibfield  {author} {\bibinfo {author} {\bibfnamefont {S.~M.}\ \bibnamefont
  {Carroll}},\ }\href@noop {} {\bibfield  {journal} {\bibinfo  {journal}
  {Living Rev.Rel.}\ }\textbf {\bibinfo {volume} {4}},\ \bibinfo {pages} {1}
  (\bibinfo {year} {2001})},\ \Eprint {http://arxiv.org/abs/astro-ph/0004075}
  {arXiv:astro-ph/0004075 [astro-ph]} \BibitemShut {NoStop}%
\bibitem [{\citenamefont {Mota}\ \emph {et~al.}(2008)\citenamefont {Mota},
  \citenamefont {Pettorino}, \citenamefont {Robbers},\ and\ \citenamefont
  {Wetterich}}]{Mota08}%
  \BibitemOpen
  \bibfield  {author} {\bibinfo {author} {\bibfnamefont {D.}~\bibnamefont
  {Mota}}, \bibinfo {author} {\bibfnamefont {V.}~\bibnamefont {Pettorino}},
  \bibinfo {author} {\bibfnamefont {G.}~\bibnamefont {Robbers}}, \ and\
  \bibinfo {author} {\bibfnamefont {C.}~\bibnamefont {Wetterich}},\ }\href
  {\doibase 10.1016/j.physletb.2008.03.060} {\bibfield  {journal} {\bibinfo
  {journal} {Phys.Lett.}\ }\textbf {\bibinfo {volume} {B663}},\ \bibinfo
  {pages} {160} (\bibinfo {year} {2008})},\ \Eprint
  {http://arxiv.org/abs/0802.1515} {arXiv:0802.1515 [astro-ph]} \BibitemShut
  {NoStop}%
\bibitem [{\citenamefont {Wetterich}(1988)}]{Wetterich88}%
  \BibitemOpen
  \bibfield  {author} {\bibinfo {author} {\bibfnamefont {C.}~\bibnamefont
  {Wetterich}},\ }\href {\doibase 10.1016/0550-3213(88)90193-9} {\bibfield
  {journal} {\bibinfo  {journal} {Nucl.Phys.}\ }\textbf {\bibinfo {volume}
  {B302}},\ \bibinfo {pages} {668} (\bibinfo {year} {1988})}\BibitemShut
  {NoStop}%
\bibitem [{\citenamefont {Ratra}\ and\ \citenamefont
  {Peebles}(1988)}]{Ratra88}%
  \BibitemOpen
  \bibfield  {author} {\bibinfo {author} {\bibfnamefont {B.}~\bibnamefont
  {Ratra}}\ and\ \bibinfo {author} {\bibfnamefont {P.}~\bibnamefont
  {Peebles}},\ }\href {\doibase 10.1103/PhysRevD.37.3406} {\bibfield  {journal}
  {\bibinfo  {journal} {Phys.Rev.}\ }\textbf {\bibinfo {volume} {D37}},\
  \bibinfo {pages} {3406} (\bibinfo {year} {1988})}\BibitemShut {NoStop}%
\bibitem [{\citenamefont {Wetterich}(2013)}]{Wetterich13}%
  \BibitemOpen
  \bibfield  {author} {\bibinfo {author} {\bibfnamefont {C.}~\bibnamefont
  {Wetterich}},\ }\href {\doibase 10.1016/j.dark.2013.10.002} {\bibfield
  {journal} {\bibinfo  {journal} {Phys.Dark Univ.}\ }\textbf {\bibinfo {volume}
  {2}},\ \bibinfo {pages} {184} (\bibinfo {year} {2013})},\ \Eprint
  {http://arxiv.org/abs/1303.6878} {arXiv:1303.6878 [astro-ph.CO]} \BibitemShut
  {NoStop}%
\bibitem [{\citenamefont {Wetterich}(2014)}]{Wetterich14}%
  \BibitemOpen
  \bibfield  {author} {\bibinfo {author} {\bibfnamefont {C.}~\bibnamefont
  {Wetterich}},\ }\href@noop {} {\  (\bibinfo {year} {2014})},\ \Eprint
  {http://arxiv.org/abs/1404.0535} {arXiv:1404.0535 [gr-qc]} \BibitemShut
  {NoStop}%
\bibitem [{\citenamefont {Pettorino}\ \emph {et~al.}(2010)\citenamefont
  {Pettorino}, \citenamefont {Wintergerst}, \citenamefont {Amendola},\ and\
  \citenamefont {Wetterich}}]{Pettorino10}%
  \BibitemOpen
  \bibfield  {author} {\bibinfo {author} {\bibfnamefont {V.}~\bibnamefont
  {Pettorino}}, \bibinfo {author} {\bibfnamefont {N.}~\bibnamefont
  {Wintergerst}}, \bibinfo {author} {\bibfnamefont {L.}~\bibnamefont
  {Amendola}}, \ and\ \bibinfo {author} {\bibfnamefont {C.}~\bibnamefont
  {Wetterich}},\ }\href {\doibase 10.1103/PhysRevD.82.123001} {\bibfield
  {journal} {\bibinfo  {journal} {Phys.Rev.}\ }\textbf {\bibinfo {volume}
  {D82}},\ \bibinfo {pages} {123001} (\bibinfo {year} {2010})},\ \Eprint
  {http://arxiv.org/abs/1009.2461} {arXiv:1009.2461 [astro-ph.CO]} \BibitemShut
  {NoStop}%
\bibitem [{\citenamefont {Ayaita}\ \emph {et~al.}(2012)\citenamefont {Ayaita},
  \citenamefont {Weber},\ and\ \citenamefont {Wetterich}}]{Ayaita12a}%
  \BibitemOpen
  \bibfield  {author} {\bibinfo {author} {\bibfnamefont {Y.}~\bibnamefont
  {Ayaita}}, \bibinfo {author} {\bibfnamefont {M.}~\bibnamefont {Weber}}, \
  and\ \bibinfo {author} {\bibfnamefont {C.}~\bibnamefont {Wetterich}},\ }\href
  {\doibase 10.1103/PhysRevD.85.123010} {\bibfield  {journal} {\bibinfo
  {journal} {Phys.Rev.}\ }\textbf {\bibinfo {volume} {D85}},\ \bibinfo {pages}
  {123010} (\bibinfo {year} {2012})},\ \Eprint {http://arxiv.org/abs/1112.4762}
  {arXiv:1112.4762 [astro-ph.CO]} \BibitemShut {NoStop}%
\bibitem [{\citenamefont {Ayaita}\ \emph {et~al.}(2013)\citenamefont {Ayaita},
  \citenamefont {Weber},\ and\ \citenamefont {Wetterich}}]{Ayaita12b}%
  \BibitemOpen
  \bibfield  {author} {\bibinfo {author} {\bibfnamefont {Y.}~\bibnamefont
  {Ayaita}}, \bibinfo {author} {\bibfnamefont {M.}~\bibnamefont {Weber}}, \
  and\ \bibinfo {author} {\bibfnamefont {C.}~\bibnamefont {Wetterich}},\ }\href
  {\doibase 10.1103/PhysRevD.87.043519} {\bibfield  {journal} {\bibinfo
  {journal} {Phys.Rev.}\ }\textbf {\bibinfo {volume} {D87}},\ \bibinfo {pages}
  {043519} (\bibinfo {year} {2013})},\ \Eprint {http://arxiv.org/abs/1211.6589}
  {arXiv:1211.6589 [astro-ph.CO]} \BibitemShut {NoStop}%
\bibitem [{\citenamefont {Fardon}\ \emph {et~al.}(2004)\citenamefont {Fardon},
  \citenamefont {Nelson},\ and\ \citenamefont {Weiner}}]{Fardon03}%
  \BibitemOpen
  \bibfield  {author} {\bibinfo {author} {\bibfnamefont {R.}~\bibnamefont
  {Fardon}}, \bibinfo {author} {\bibfnamefont {A.~E.}\ \bibnamefont {Nelson}},
  \ and\ \bibinfo {author} {\bibfnamefont {N.}~\bibnamefont {Weiner}},\ }\href
  {\doibase 10.1088/1475-7516/2004/10/005} {\bibfield  {journal} {\bibinfo
  {journal} {JCAP}\ }\textbf {\bibinfo {volume} {0410}},\ \bibinfo {pages}
  {005} (\bibinfo {year} {2004})},\ \Eprint
  {http://arxiv.org/abs/astro-ph/0309800} {arXiv:astro-ph/0309800 [astro-ph]}
  \BibitemShut {NoStop}%
\bibitem [{\citenamefont {Afshordi}\ \emph {et~al.}(2005)\citenamefont
  {Afshordi}, \citenamefont {Zaldarriaga},\ and\ \citenamefont
  {Kohri}}]{Afshordi05}%
  \BibitemOpen
  \bibfield  {author} {\bibinfo {author} {\bibfnamefont {N.}~\bibnamefont
  {Afshordi}}, \bibinfo {author} {\bibfnamefont {M.}~\bibnamefont
  {Zaldarriaga}}, \ and\ \bibinfo {author} {\bibfnamefont {K.}~\bibnamefont
  {Kohri}},\ }\href {\doibase 10.1103/PhysRevD.72.065024} {\bibfield  {journal}
  {\bibinfo  {journal} {Phys.Rev.}\ }\textbf {\bibinfo {volume} {D72}},\
  \bibinfo {pages} {065024} (\bibinfo {year} {2005})},\ \Eprint
  {http://arxiv.org/abs/astro-ph/0506663} {arXiv:astro-ph/0506663 [astro-ph]}
  \BibitemShut {NoStop}%
\bibitem [{\citenamefont {Bjaelde}\ \emph {et~al.}(2008)\citenamefont
  {Bjaelde}, \citenamefont {Brookfield}, \citenamefont {van~de Bruck},
  \citenamefont {Hannestad}, \citenamefont {Mota} \emph {et~al.}}]{Bjaelde07}%
  \BibitemOpen
  \bibfield  {author} {\bibinfo {author} {\bibfnamefont {O.~E.}\ \bibnamefont
  {Bjaelde}}, \bibinfo {author} {\bibfnamefont {A.~W.}\ \bibnamefont
  {Brookfield}}, \bibinfo {author} {\bibfnamefont {C.}~\bibnamefont {van~de
  Bruck}}, \bibinfo {author} {\bibfnamefont {S.}~\bibnamefont {Hannestad}},
  \bibinfo {author} {\bibfnamefont {D.~F.}\ \bibnamefont {Mota}},  \emph
  {et~al.},\ }\href {\doibase 10.1088/1475-7516/2008/01/026} {\bibfield
  {journal} {\bibinfo  {journal} {JCAP}\ }\textbf {\bibinfo {volume} {0801}},\
  \bibinfo {pages} {026} (\bibinfo {year} {2008})},\ \Eprint
  {http://arxiv.org/abs/0705.2018} {arXiv:0705.2018 [astro-ph]} \BibitemShut
  {NoStop}%
\bibitem [{\citenamefont {Brouzakis}\ \emph {et~al.}(2008)\citenamefont
  {Brouzakis}, \citenamefont {Tetradis},\ and\ \citenamefont
  {Wetterich}}]{Brouzakis07}%
  \BibitemOpen
  \bibfield  {author} {\bibinfo {author} {\bibfnamefont {N.}~\bibnamefont
  {Brouzakis}}, \bibinfo {author} {\bibfnamefont {N.}~\bibnamefont {Tetradis}},
  \ and\ \bibinfo {author} {\bibfnamefont {C.}~\bibnamefont {Wetterich}},\
  }\href {\doibase 10.1016/j.physletb.2008.05.068} {\bibfield  {journal}
  {\bibinfo  {journal} {Phys.Lett.}\ }\textbf {\bibinfo {volume} {B665}},\
  \bibinfo {pages} {131} (\bibinfo {year} {2008})},\ \Eprint
  {http://arxiv.org/abs/0711.2226} {arXiv:0711.2226 [astro-ph]} \BibitemShut
  {NoStop}%
\bibitem [{\citenamefont {Wetterich}(2008)}]{Wetterich08}%
  \BibitemOpen
  \bibfield  {author} {\bibinfo {author} {\bibfnamefont {C.}~\bibnamefont
  {Wetterich}},\ }\href {\doibase 10.1103/PhysRevD.77.103505} {\bibfield
  {journal} {\bibinfo  {journal} {Phys.Rev.}\ }\textbf {\bibinfo {volume}
  {D77}},\ \bibinfo {pages} {103505} (\bibinfo {year} {2008})},\ \Eprint
  {http://arxiv.org/abs/0801.3208} {arXiv:0801.3208 [hep-th]} \BibitemShut
  {NoStop}%
\bibitem [{\citenamefont {Doran}\ \emph {et~al.}(2007)\citenamefont {Doran},
  \citenamefont {Robbers},\ and\ \citenamefont {Wetterich}}]{Doran07}%
  \BibitemOpen
  \bibfield  {author} {\bibinfo {author} {\bibfnamefont {M.}~\bibnamefont
  {Doran}}, \bibinfo {author} {\bibfnamefont {G.}~\bibnamefont {Robbers}}, \
  and\ \bibinfo {author} {\bibfnamefont {C.}~\bibnamefont {Wetterich}},\ }\href
  {\doibase 10.1103/PhysRevD.75.023003} {\bibfield  {journal} {\bibinfo
  {journal} {Phys.Rev.}\ }\textbf {\bibinfo {volume} {D75}},\ \bibinfo {pages}
  {023003} (\bibinfo {year} {2007})},\ \Eprint
  {http://arxiv.org/abs/astro-ph/0609814} {arXiv:astro-ph/0609814 [astro-ph]}
  \BibitemShut {NoStop}%
\bibitem [{\citenamefont {Reichardt}\ \emph {et~al.}(2012)\citenamefont
  {Reichardt}, \citenamefont {de~Putter}, \citenamefont {Zahn},\ and\
  \citenamefont {Hou}}]{Reichardt12}%
  \BibitemOpen
  \bibfield  {author} {\bibinfo {author} {\bibfnamefont {C.~L.}\ \bibnamefont
  {Reichardt}}, \bibinfo {author} {\bibfnamefont {R.}~\bibnamefont
  {de~Putter}}, \bibinfo {author} {\bibfnamefont {O.}~\bibnamefont {Zahn}}, \
  and\ \bibinfo {author} {\bibfnamefont {Z.}~\bibnamefont {Hou}},\ }\href
  {\doibase 10.1088/2041-8205/749/1/L9} {\bibfield  {journal} {\bibinfo
  {journal} {Astrophys.J.}\ }\textbf {\bibinfo {volume} {749}},\ \bibinfo
  {pages} {L9} (\bibinfo {year} {2012})},\ \Eprint
  {http://arxiv.org/abs/1110.5328} {arXiv:1110.5328 [astro-ph.CO]} \BibitemShut
  {NoStop}%
\bibitem [{\citenamefont {Pettorino}\ \emph {et~al.}(2013)\citenamefont
  {Pettorino}, \citenamefont {Amendola},\ and\ \citenamefont
  {Wetterich}}]{Pettorino13}%
  \BibitemOpen
  \bibfield  {author} {\bibinfo {author} {\bibfnamefont {V.}~\bibnamefont
  {Pettorino}}, \bibinfo {author} {\bibfnamefont {L.}~\bibnamefont {Amendola}},
  \ and\ \bibinfo {author} {\bibfnamefont {C.}~\bibnamefont {Wetterich}},\
  }\href {\doibase 10.1103/PhysRevD.87.083009} {\bibfield  {journal} {\bibinfo
  {journal} {Phys.Rev.}\ }\textbf {\bibinfo {volume} {D87}},\ \bibinfo {pages}
  {083009} (\bibinfo {year} {2013})},\ \Eprint {http://arxiv.org/abs/1301.5279}
  {arXiv:1301.5279 [astro-ph.CO]} \BibitemShut {NoStop}%
\bibitem [{\citenamefont {Magg}\ and\ \citenamefont
  {Wetterich}(1980)}]{Magg80}%
  \BibitemOpen
  \bibfield  {author} {\bibinfo {author} {\bibfnamefont {M.}~\bibnamefont
  {Magg}}\ and\ \bibinfo {author} {\bibfnamefont {C.}~\bibnamefont
  {Wetterich}},\ }\href {\doibase 10.1016/0370-2693(80)90825-4} {\bibfield
  {journal} {\bibinfo  {journal} {Phys.Lett.}\ }\textbf {\bibinfo {volume}
  {B94}},\ \bibinfo {pages} {61} (\bibinfo {year} {1980})}\BibitemShut
  {NoStop}%
\bibitem [{\citenamefont {Lazarides}\ \emph {et~al.}(1981)\citenamefont
  {Lazarides}, \citenamefont {Shafi},\ and\ \citenamefont
  {Wetterich}}]{Lazarides81}%
  \BibitemOpen
  \bibfield  {author} {\bibinfo {author} {\bibfnamefont {G.}~\bibnamefont
  {Lazarides}}, \bibinfo {author} {\bibfnamefont {Q.}~\bibnamefont {Shafi}}, \
  and\ \bibinfo {author} {\bibfnamefont {C.}~\bibnamefont {Wetterich}},\ }\href
  {\doibase 10.1016/0550-3213(81)90354-0} {\bibfield  {journal} {\bibinfo
  {journal} {Nucl.Phys.}\ }\textbf {\bibinfo {volume} {B181}},\ \bibinfo
  {pages} {287} (\bibinfo {year} {1981})}\BibitemShut {NoStop}%
\bibitem [{\citenamefont {Mohapatra}\ and\ \citenamefont
  {Senjanovic}(1981)}]{Mohapatra80}%
  \BibitemOpen
  \bibfield  {author} {\bibinfo {author} {\bibfnamefont {R.~N.}\ \bibnamefont
  {Mohapatra}}\ and\ \bibinfo {author} {\bibfnamefont {G.}~\bibnamefont
  {Senjanovic}},\ }\href {\doibase 10.1103/PhysRevD.23.165} {\bibfield
  {journal} {\bibinfo  {journal} {Phys.Rev.}\ }\textbf {\bibinfo {volume}
  {D23}},\ \bibinfo {pages} {165} (\bibinfo {year} {1981})}\BibitemShut
  {NoStop}%
\bibitem [{\citenamefont {Schechter}\ and\ \citenamefont
  {Valle}(1980)}]{Schechter80}%
  \BibitemOpen
  \bibfield  {author} {\bibinfo {author} {\bibfnamefont {J.}~\bibnamefont
  {Schechter}}\ and\ \bibinfo {author} {\bibfnamefont {J.}~\bibnamefont
  {Valle}},\ }\href {\doibase 10.1103/PhysRevD.22.2227} {\bibfield  {journal}
  {\bibinfo  {journal} {Phys.Rev.}\ }\textbf {\bibinfo {volume} {D22}},\
  \bibinfo {pages} {2227} (\bibinfo {year} {1980})}\BibitemShut {NoStop}%
\bibitem [{\citenamefont {Wetterich}(1995)}]{Wetterich95}%
  \BibitemOpen
  \bibfield  {author} {\bibinfo {author} {\bibfnamefont {C.}~\bibnamefont
  {Wetterich}},\ }\href@noop {} {\bibfield  {journal} {\bibinfo  {journal}
  {Astron.Astrophys.}\ }\textbf {\bibinfo {volume} {301}},\ \bibinfo {pages}
  {321} (\bibinfo {year} {1995})},\ \Eprint
  {http://arxiv.org/abs/hep-th/9408025} {arXiv:hep-th/9408025 [hep-th]}
  \BibitemShut {NoStop}%
\bibitem [{\citenamefont {Amendola}(2000)}]{Amendola00}%
  \BibitemOpen
  \bibfield  {author} {\bibinfo {author} {\bibfnamefont {L.}~\bibnamefont
  {Amendola}},\ }\href {\doibase 10.1103/PhysRevD.62.043511} {\bibfield
  {journal} {\bibinfo  {journal} {Phys.Rev.}\ }\textbf {\bibinfo {volume}
  {D62}},\ \bibinfo {pages} {043511} (\bibinfo {year} {2000})},\ \Eprint
  {http://arxiv.org/abs/astro-ph/9908023} {arXiv:astro-ph/9908023 [astro-ph]}
  \BibitemShut {NoStop}%
\bibitem [{\citenamefont {Wintergerst}\ \emph {et~al.}(2010)\citenamefont
  {Wintergerst}, \citenamefont {Pettorino}, \citenamefont {Mota},\ and\
  \citenamefont {Wetterich}}]{Wintergerst10a}%
  \BibitemOpen
  \bibfield  {author} {\bibinfo {author} {\bibfnamefont {N.}~\bibnamefont
  {Wintergerst}}, \bibinfo {author} {\bibfnamefont {V.}~\bibnamefont
  {Pettorino}}, \bibinfo {author} {\bibfnamefont {D.}~\bibnamefont {Mota}}, \
  and\ \bibinfo {author} {\bibfnamefont {C.}~\bibnamefont {Wetterich}},\ }\href
  {\doibase 10.1103/PhysRevD.81.063525} {\bibfield  {journal} {\bibinfo
  {journal} {Phys.Rev.}\ }\textbf {\bibinfo {volume} {D81}},\ \bibinfo {pages}
  {063525} (\bibinfo {year} {2010})},\ \Eprint {http://arxiv.org/abs/0910.4985}
  {arXiv:0910.4985 [astro-ph.CO]} \BibitemShut {NoStop}%
\bibitem [{\citenamefont {Schrempp}\ and\ \citenamefont
  {Brown}(2010)}]{Schrempp09}%
  \BibitemOpen
  \bibfield  {author} {\bibinfo {author} {\bibfnamefont {L.}~\bibnamefont
  {Schrempp}}\ and\ \bibinfo {author} {\bibfnamefont {I.}~\bibnamefont
  {Brown}},\ }\href {\doibase 10.1088/1475-7516/2010/05/023} {\bibfield
  {journal} {\bibinfo  {journal} {JCAP}\ }\textbf {\bibinfo {volume} {1005}},\
  \bibinfo {pages} {023} (\bibinfo {year} {2010})},\ \Eprint
  {http://arxiv.org/abs/0912.3157} {arXiv:0912.3157 [astro-ph.CO]} \BibitemShut
  {NoStop}%
\bibitem [{\citenamefont {Nunes}\ \emph {et~al.}(2011)\citenamefont {Nunes},
  \citenamefont {Schrempp},\ and\ \citenamefont {Wetterich}}]{Nunes11}%
  \BibitemOpen
  \bibfield  {author} {\bibinfo {author} {\bibfnamefont {N.~J.}\ \bibnamefont
  {Nunes}}, \bibinfo {author} {\bibfnamefont {L.}~\bibnamefont {Schrempp}}, \
  and\ \bibinfo {author} {\bibfnamefont {C.}~\bibnamefont {Wetterich}},\ }\href
  {\doibase 10.1103/PhysRevD.83.083523} {\bibfield  {journal} {\bibinfo
  {journal} {Phys.Rev.}\ }\textbf {\bibinfo {volume} {D83}},\ \bibinfo {pages}
  {083523} (\bibinfo {year} {2011})},\ \Eprint {http://arxiv.org/abs/1102.1664}
  {arXiv:1102.1664 [astro-ph.CO]} \BibitemShut {NoStop}%
\bibitem [{\citenamefont {Baldi}\ \emph {et~al.}(2011)\citenamefont {Baldi},
  \citenamefont {Pettorino}, \citenamefont {Amendola},\ and\ \citenamefont
  {Wetterich}}]{Baldi11a}%
  \BibitemOpen
  \bibfield  {author} {\bibinfo {author} {\bibfnamefont {M.}~\bibnamefont
  {Baldi}}, \bibinfo {author} {\bibfnamefont {V.}~\bibnamefont {Pettorino}},
  \bibinfo {author} {\bibfnamefont {L.}~\bibnamefont {Amendola}}, \ and\
  \bibinfo {author} {\bibfnamefont {C.}~\bibnamefont {Wetterich}},\ }\href@noop
  {} {\  (\bibinfo {year} {2011})},\ \Eprint {http://arxiv.org/abs/1106.2161}
  {arXiv:1106.2161 [astro-ph.CO]} \BibitemShut {NoStop}%
\bibitem [{\citenamefont {Ayaita}\ \emph {et~al.}(2009)\citenamefont {Ayaita},
  \citenamefont {Weber},\ and\ \citenamefont {Wetterich}}]{Ayaita09}%
  \BibitemOpen
  \bibfield  {author} {\bibinfo {author} {\bibfnamefont {Y.}~\bibnamefont
  {Ayaita}}, \bibinfo {author} {\bibfnamefont {M.}~\bibnamefont {Weber}}, \
  and\ \bibinfo {author} {\bibfnamefont {C.}~\bibnamefont {Wetterich}},\
  }\href@noop {} {\  (\bibinfo {year} {2009})},\ \Eprint
  {http://arxiv.org/abs/0908.2903} {arXiv:0908.2903 [astro-ph.CO]} \BibitemShut
  {NoStop}%
\bibitem [{\citenamefont {Puchwein}\ \emph {et~al.}(2013)\citenamefont
  {Puchwein}, \citenamefont {Baldi},\ and\ \citenamefont
  {Springel}}]{Puchwein13}%
  \BibitemOpen
  \bibfield  {author} {\bibinfo {author} {\bibfnamefont {E.}~\bibnamefont
  {Puchwein}}, \bibinfo {author} {\bibfnamefont {M.}~\bibnamefont {Baldi}}, \
  and\ \bibinfo {author} {\bibfnamefont {V.}~\bibnamefont {Springel}},\
  }\href@noop {} {\bibfield  {journal} {\bibinfo  {journal}
  {Mon.Not.Roy.Astron.Soc.}\ }\textbf {\bibinfo {volume} {436}},\ \bibinfo
  {pages} {348} (\bibinfo {year} {2013})},\ \Eprint
  {http://arxiv.org/abs/1305.2418} {arXiv:1305.2418 [astro-ph.CO]} \BibitemShut
  {NoStop}%
\bibitem [{\citenamefont {Oyaizu}(2008)}]{Oyaizu08}%
  \BibitemOpen
  \bibfield  {author} {\bibinfo {author} {\bibfnamefont {H.}~\bibnamefont
  {Oyaizu}},\ }\href {\doibase 10.1103/PhysRevD.78.123523} {\bibfield
  {journal} {\bibinfo  {journal} {Phys.Rev.}\ }\textbf {\bibinfo {volume}
  {D78}},\ \bibinfo {pages} {123523} (\bibinfo {year} {2008})},\ \Eprint
  {http://arxiv.org/abs/0807.2449} {arXiv:0807.2449 [astro-ph]} \BibitemShut
  {NoStop}%
\bibitem [{\citenamefont {Wintergerst}\ and\ \citenamefont
  {Pettorino}(2010)}]{Wintergerst10b}%
  \BibitemOpen
  \bibfield  {author} {\bibinfo {author} {\bibfnamefont {N.}~\bibnamefont
  {Wintergerst}}\ and\ \bibinfo {author} {\bibfnamefont {V.}~\bibnamefont
  {Pettorino}},\ }\href {\doibase 10.1103/PhysRevD.82.103516} {\bibfield
  {journal} {\bibinfo  {journal} {Phys.Rev.}\ }\textbf {\bibinfo {volume}
  {D82}},\ \bibinfo {pages} {103516} (\bibinfo {year} {2010})},\ \Eprint
  {http://arxiv.org/abs/1005.1278} {arXiv:1005.1278 [astro-ph.CO]} \BibitemShut
  {NoStop}%
\end{thebibliography}%

\end{document}